\documentclass[12pt,a4paper]{article}
\usepackage[utf8]{inputenc}
\usepackage[english]{babel}
\usepackage[left=2.5cm,right=2.5cm,top=2.4cm,bottom=2.5cm]{geometry}
\usepackage{amsmath,amssymb,amsfonts}
\usepackage{amsthm}
\usepackage[toc,page]{appendix}
\usepackage{setspace}
\usepackage[skip=10pt, font=small, labelfont=bf]{caption}
\usepackage{mdframed}
\usepackage{enumitem}
\usepackage{ragged2e}
\usepackage{algorithm}
\usepackage{algorithmic}
\usepackage{multirow}
\usepackage{longtable}
\usepackage{rotating}
\usepackage{float}
\usepackage[colorlinks,citecolor=blue,linkcolor=blue,urlcolor=blue,pagebackref]{hyperref}
\onehalfspace 
\usepackage{natbib}

\newtheorem{assumpt}{Assumption}
\newtheorem{theo}{Theorem}

\newtheorem{example}{Example}
\newtheorem{defi}{Definition}

\newcommand{\bb}[1]{\mathbf{#1}}

\DeclareMathOperator{\Y}{Y}
\DeclareMathOperator{\X}{X}
\DeclareMathOperator{\W}{W}
\DeclareMathOperator{\E}{E}
\DeclareMathOperator*{\argmin}{argmin}

\makeatletter
\newenvironment{MyBox}[2]{ 
\protected@edef\@currentlabelname{#2}
\protected@edef\@currentlabel{#2}
\begin{mdframed}[
innerlinewidth=0.5pt,
innerleftmargin=10pt,
innerrightmargin=10pt,
innertopmargin = 10pt,
innerbottommargin=10pt,
frametitle={#1},
frametitlerule=true,
frametitlerulewidth=1pt]
}{
\end{mdframed}
}
\makeatother

\begin{document}

\author{
Augusto Cerqua$^\dagger$, Marco Letta$^\dagger$, Fiammetta Menchetti$^\ddagger$  
}
\title{Causal inference and policy evaluation \\ without a control group$^{*}$}

\date{First version: December 2022 \\This version: October 2024}

\maketitle
\begin{abstract}
\frenchspacing Without a control group, the most widespread methodologies for estimating causal effects cannot be applied. To fill this gap, we propose the Machine Learning Control Method, a new approach for causal panel analysis that estimates causal parameters without relying on untreated units. We formalize identification within the potential outcomes framework and then provide estimation based on machine learning algorithms. To illustrate the practical relevance of our method, we present simulation evidence, a replication study, and an empirical application on the impact of the COVID-19 crisis on educational inequality. We implement the proposed approach in the companion \textsf{R} package \texttt{MachineControl}.
\end{abstract}


\textbf{JEL-Codes:} C18, C21, C53, I24.

\vspace{0pt}
---------------------
\begin{spacing}{1}
\begin{footnotesize}
$^\dagger$ Department of Social Sciences and Economics, Sapienza University of Rome, Rome, Italy (IT). Email at: \texttt{augusto.cerqua@uniroma1.it}; \texttt{marco.letta@uniroma1.it} 
\end{footnotesize}

\begin{footnotesize}
$^\ddagger$ Department of Statistics, Computer Science and Applications, University of Florence, Florence, Italy (IT). Email at: \texttt{fiammetta.menchetti@unifi.it}
\end{footnotesize}


\vspace{5pt}
\begin{footnotesize}
*We are grateful to Guido Imbens and Fabrizia Mealli for thoughtful conversations and discussions. We also thank Andrea Albanese, Guglielmo Barone, Michele Battisti, Iavor Bojinov, Fabrizio Cipollini, Alessio D’Ignazio, Christina Gatmann, Anna Gottard, Giulio Grossi, Martin Huber, Michael Knaus, Joanna Kopinska, Giuseppe Maggio, Alessandra Mattei, Raffaele Mattera, Giovanni Mellace, Andrea Mercatanti, Samuel Nocito, Gabriele Pinto, Jason Poulos, Donato Romano, Federico Rucci, Jacques-François Thisse, Luca Tiberti, Giuseppe Ragusa, Giuliano Resce, and Bas van der Klaauw for valuable suggestions on earlier versions of this work. The paper has benefited from helpful comments and suggestions by audiences at many conferences and seminars.
\end{footnotesize}
\end{spacing}
\newpage
\newgeometry{left=2.25cm,right=2.25cm,top=2.5cm,bottom=2.5cm}
\begin{flushright}
\textit{``In history there are no control groups. There is no one to tell us what might have been.''}\\Cormac McCarthy, All the Pretty Horses (1992)
\end{flushright} 
\vspace{5pt}
\section{Introduction}
\label{sec:intro}
\frenchspacing Today, the econometric toolbox for causal panel analysis features many alternative approaches for estimating causal effects, such as difference-in-differences \citep{Card:Krueger:1994}, the synthetic control method   \citep{Abadie:Diamond:Hainmueller:2010}, two-way \citep{Angrist:Pischke:2009} and interactive \citep{Bai:2009} fixed effects models, and matrix completion methods \citep{Athey:Bayati:Doudchenko:Imbens:Khosravi:2021}. However, all of these popular methodologies depend on a critical requirement: the availability of untreated units forming a credible control group, without which they cannot be applied. This poses a significant empirical challenge in observational studies, as there are at least two relevant cases in which a suitable control group does not exist: (i) when the treatment simultaneously affects all units, such as in the context of a large-scale shock or a nationwide program with universal participation \citep{Duflo:2017}; (ii) when only a subgroup of units receives the treatment, but the set of untreated units cannot form a valid control group due to violations of the no-interference assumption \citep{Cox:1958}.\footnote{Spillovers are common in many empirical settings where the observations are correlated in time, space, or both, and ignoring them can lead to very misleading inferences \citep{Sobel:2006}. In principle, spillovers can be modeled and accounted for at the cost of additional and often restrictive assumptions. However, in practice, most methods just postulate the absence of interference.} Under these challenging but not uncommon circumstances, the use of standard causal panel data methods is precluded, leaving a gap in the econometric toolkit of empirical researchers.

\frenchspacing We fill this gap by introducing the Machine Learning Control Method (MLCM), a new estimator based on flexible counterfactual forecasting via machine learning (ML). The MLCM is a versatile technique that can leverage any available supervised ML algorithm and enables the identification and estimation of several policy-relevant causal parameters---including individual, average, and conditional average treatment effects (CATEs)---in practically relevant environments without a control group. It is flexible in terms of data structure, as it can be applied across a wide variety of panel settings, including very short panels and contexts with staggered adoption. The intuition behind the MLCM is straightforward: when you cannot rely on untreated units to build the counterfactual, you can forecast it. Specifically, use supervised ML techniques trained and tuned on the pre-treatment data to forecast the evolution of post-treatment outcomes in the absence of the treatment. Then, estimate treatment effects as the difference between observed and forecasted post-treatment outcomes.

Our approach builds upon and bridges three different methodological currents: causal panel data methods, time series forecasting, and causal ML. From the literature on causal panel analysis, we adopt the conceptual and identification framework based on the potential outcomes model \citep{Rubin:1974}. Since earlier contributions \citep{Ashenfelter:Card:1985, Card:1990} to its more recent developments \citep{Arkhangelsky:Athey:Hirshberg:Imbens:Wager:2021, Roth:Santanna:Bilinski:Poe:2023}, this entire literature revolves around addressing the ``fundamental problem of causal inference'' \citep{Holland:1986} in an observational panel data setting where, for each unit, it is impossible to observe both potential outcomes at once. Therefore, causal inference can be viewed as a missing data problem, with the key goal being the imputation of missing potential outcomes for the treated units \citep{Imbens:Rubin:2015}. Most widely used causal panel data methods achieve this imputation by leveraging untreated units to construct a control group. Our key contribution to this literature is to establish identification conditions in the absence of a control group and to introduce a novel estimator of causal effects that imputes missing potential outcomes without relying on untreated units.

The time series literature features some forecasting approaches that researchers have employed for counterfactual building in scenarios without a control group.\footnote{A rich tradition also exists for forecasting with panel data (e.g., \citet{arellano2011nonlinear, baltagi2008econometric, liu2020forecasting}). However, our interest lies in \textit{counterfactual} forecasting. In this regard, the Rubin causal framework is absent from this tradition, which focuses on different estimands and non-causal settings compared to the recent causal panel data literature \citep{arkhangelsky2024causal}.} Some of them are straightforward: the mean method estimates the counterfactual ``no-treatment scenario'' of the dependent variable $\Y_{i}$ as the average of its pre-treatment values, while the naive method considers the last pre-treatment value of $\Y_{i}$ as the counterfactual \citep{Hyndman:Athanasopoulos:2021}. A more sophisticated forecasting approach is the interrupted time series (ITS) analysis, first introduced by \citet{Box:Tiao:1975}. ITS analysis generally fits regression models or autoregressive integrated moving average (ARIMA) models to the entire time series of observed data; this is done after postulating a structure on the intervention effect (e.g., constant level shift). By construction, ITS delivers a single average effect for the whole post-intervention period. More recent studies \citep{Brodersen:Gallusser:Koehler:Remy:Scott:2015, Chernozhukov:Wuthrich:Zhu:2021,Menchetti:Cipollini:Mealli:2023, rambachan2021common} consider estimation and inference with other time series techniques (such as Bayesian Structural Time Series, ARIMA models, or impulse response functions) in time series settings where no direct control group may be available. From a panel data perspective, these methods have significant limitations. First, they are designed for cases with a single treated series. Second, they often impose restrictive assumptions and functional forms. Third, they can be implemented only when many pre-treatment periods are available. However, in the traditional panel data literature, the number of units is much larger than the number of time periods \citep{arkhangelsky2024causal}, and the time dimension is typically substantially smaller than in time series settings, rendering these methods mostly inapplicable. We draw from this literature the idea of forecasting counterfactuals by exploiting the temporal information. We go beyond it by proposing a more flexible, ML-powered method explicitly designed for counterfactual forecasting with panel data.

In recent years, a new methodological literature at the intersection between causal inference with panel data and ML has evolved. The matrix completion methods developed by \citet{Athey:Bayati:Doudchenko:Imbens:Khosravi:2021} use nuclear norm regularization and observed control outcomes to impute missing elements in the counterfactual matrix of untreated unit/period combinations. \citet{semenova2017estimation} propose an approach for estimating CATEs characterized by high-dimensional parameters in both homogeneous cross-sectional and unit-heterogeneous dynamic panel data settings, leveraging ML techniques for improved inference. Artificial control methods \citep{Carvalho:Masini:Medeiros:2018,Masini:Medeiros:2021} and synthetic learners \citep{Viviano:Bradic:2023} harness supervised ML algorithms to predict counterfactuals and assess treatment effects in a panel setting with a single treated unit, many untreated units, and a long pre-intervention window. All these methods need a set of units assumed to be completely unaffected by the treatment to build a counterfactual scenario. Therefore, none of these causal ML techniques can be applied in the challenging econometric setting we focus on. We share with this literature the idea of harnessing the power of ML in the service of causal inference rather than for pure prediction, exploiting the fact that counterfactual building is ultimately a predictive task \citep{Varian:2016}. We extend it by developing a new causal ML estimator that estimates individual treatment effects without depending on the no-interference assumption or leveraging unaffected units. 

We begin by thoroughly discussing and formalizing identification without a control group in the potential outcomes framework. We then propose an adaptive estimator based on a horse-race competition between several different ML algorithms, a novel cross-validation (CV) procedure for model assessment and selection tailored for panel data, and block-bootstrapping for inference. The MLCM comes with a set of diagnostic and placebo tests and is characterized by a high level of generality: it delivers individual treatment effects that can either be the object of interest or aggregated into several policy-relevant causal estimands. Since our approach allows for unrestricted treatment effect heterogeneity, we also propose an automated search for heterogeneity based on an easy-to-interpret regression tree. 

To showcase the  potential of the MLCM, we present an extensive simulation study, a replication of the minimum wage application in \citet{callaway2021difference} without using control units, and an empirical application in which we investigate the effects of the COVID-19 pandemic on educational inequality in Italian Local Labor Markets (LLMs). We find that the pandemic led to a generalized drop in students’ performance, which is particularly pronounced in LLMs that before COVID-19 were characterized by higher levels of unemployment and inequality and lower educational attainment.

At the time of writing, we are aware of only one other method for assessing causality in panel settings without using control units.\footnote{There are also some empirical studies in energy economics that, without proposing a new formal method, have applied counterfactual prediction using ML algorithms on very high-frequency (hourly) electricity data \citep{abrell2022effective, jarvis2022private, prest2023rcts}.} In independent work subsequent to ours, \citet{Botosaru:Giacomini:Weidner:2023} propose estimating average treatment effects in the absence of a control group by regressing pre-treatment outcomes on known basis functions of time. Compared to the MLCM, this method relies on different assumptions, such as absence of interference and the validity of a central limit theorem, and imposes functional form restrictions. Overall, we view the two methods as complementary, depending on the assumptions one is willing to make and on the characteristics of the available data.\footnote{For instance, the estimator by \citet{Botosaru:Giacomini:Weidner:2023} can be employed in data-scarce environments where only outcome data are available, or when the researcher is comfortable relying on parametric assumptions. In contrast, the MLCM is more flexible and better suited for data-rich or high-dimensional settings where information on many covariates is available. This context is particularly relevant when the primary goal is to uncover heterogeneity in treatment effects associated with differences in observed characteristics.}

The possibility of moving beyond the reliance on untreated units paves the way to evaluating many potential treatments---such as universal policies, large-scale shocks, and programs engendering interactions between units---that, due to the lack of a valid comparison group, have so far been under-explored in empirical studies. To answer these causal questions, interested researchers can implement the MLCM on real-world datasets using the companion \textsf{R} package \texttt{MachineControl}.\footnote{The latest version of the package is available on GitHub at \href{https://github.com/FMenchetti/MachineControl}{\textcolor{blue}{this link}}.}

The rest of this paper is organized as follows. Section \ref{sec:causal_framework} outlines the causal framework by defining the identification assumptions, the causal estimands, and the estimators. Section \ref{sec:implementation} describes the implementation process of the MLCM and presents the simulation study. Section \ref{sec:application} illustrates the empirical application, while Section \ref{sec:conclusion} concludes.

\section{The causal framework}
\label{sec:causal_framework}
In this section, we present the causal framework for an observational short-panel setup where the intervention is a simultaneous policy change or shock that affects directly or indirectly the entire statistical population under scrutiny.\footnote{To use language from \citet{arkhangelsky2024causal}, in this benchmark case, we consider a thin panel matrix, with a large number of cross-sectional units and a small number of time periods. However, the framework easily accommodates the use of long panels and fat and square matrices. In addition, the MLCM can be applied to both balanced and unbalanced panels, as well as in block-assignment and staggered adoption settings where one cannot credibly rely on the plausibility of the no-interference assumption (i.e., outcomes of never-treated or not-yet-treated units are contaminated by spillovers).} We start by defining the assumptions and the corresponding causal estimands within the potential outcomes framework. We then provide the proof that such estimands can be identified under those assumptions. Finally, we introduce novel estimators for scenarios without a suitable control group. To make the notation (and the method) as general as possible, the definitions given in this section assume the presence of some contemporaneous covariates unaffected by the intervention. Including these covariates in the MLCM can potentially mitigate certain assumptions. However, as we will discuss later, caution must be exercised in their use and selection.

\subsection{Assumptions}
\label{subsec:causal_assumpt}
Denote with $\Y_{i,t}$ the outcome of unit $i$ at time $t$, and let $\W_{i,t} \in \{0,1\}$ be a random variable describing the treatment assignment of unit $i = 1, \dots, N$ at time $t = 1, \dots,t_0,\dots, T$, where $1$ indicates the treatment, $0$ indicates control, and $t_0$ denotes the intervention date.\footnote{The word ``treatment'' is commonly used in the context of randomized controlled trials. As we are dealing with an observational study, we use the terms  ``treatment'', ``intervention'', ``policy”, and ``shock'' interchangeably.} As we are focusing on a single and simultaneous intervention, we can then write $\W_{i,t} = 0$ for all $i$ and $t \leq t_0$, $\W_{i,t} = 1$ for all $i$ and $t > t_0$, therefore, $t_0+1$ is the first post-intervention period. In other words, all units are treated simultaneously and remain treated thereafter. Under the potential outcomes framework, for each unit, we can define a potential outcome under $\W_{i,t} = 0$ and a different potential outcome under $\W_{i,t} = 1$. 

We now introduce the assumptions that are needed to define, identify, and estimate causal effects in an observational panel setting without controls, offering novel theoretical insights such as identification conditions for non-linear, multi-step-ahead causal effects. We emphasize that such assumptions are not particularly restrictive, as the MLCM is designed to be highly flexible.
Our method, whose implementation is detailed in Section \ref{sec:implementation}, starts by running several ML algorithms on pre-treatment data and then selects the one producing the most accurate forecasts. Given that it is not possible to know in advance which ML method will be chosen, the notation is kept as general as possible.\footnote{We believe this is an advantage over most existing approaches in the causal ML literature: while \citet{Carvalho:Masini:Medeiros:2018} and \citet{Masini:Medeiros:2021} focus on LASSO and \citet{Wager:Athey:2018} adopt tree-based methods, the MLCM is flexible and can easily adapt to different data-generating processes. A notable exception in this context is the synthetic learner proposed by \citet{Viviano:Bradic:2023}. In contrast to their technique, our method is based on a horse-race competition between several alternative ML algorithms, whereas the synthetic learner relies on an ensemble procedure that combines various estimators.
} 

The first assumption contributes to define potential outcomes and establishes the crucial link between them and the treatment assignment. We rely on a less stringent version of the Stable Unit Treatment Value Assumption (SUTVA) \citep{Rubin:1974, Imbens:Rubin:2015}, retaining only its second part: the treatment is the same for all the units. This is an \textit{a priori} assumption that the potential outcomes do not depend on different treatment intensities or the mechanism used to assign the treatment.

\begin{assumpt} [Weak SUTVA]
\label{assumpt:sutva}
There are no hidden forms of treatment leading to different potential outcomes.
\end{assumpt}

While we maintain this no-multiple-versions-of-treatment assumption, we drop the first part of SUTVA, namely, the no-interference assumption \citep{Cox:1958}.\footnote{Following \citet{Imbens:Rubin:2015}, we consider the case with general equilibrium effects as a scenario in which there are widespread violations of the no-interference assumption.} We consider this as one of the main advantages of our approach because, while it has long been known that violations and failures of the no-interference assumption can lead to misleading inferences in many social science settings \citep{Sobel:2006}, this strong assumption is rarely questioned or tested in practice \citep{Chiu:Lan:Liu:Ziyi:Xu:2023}. This is a key departure from most evaluation methods and it is important to delve into its implications. SUTVA-related interference among units can be of two types: (1) from treated to control units, and (2) among treated units themselves.\footnote{Even if the treatment is the same for all units, residual spillover effects may arise due to interference among the treated units. Consider, for example, a study investigating excess mortality from COVID-19. While the pandemic affects all areas, those with fewer intensive care beds may need to send patients to neighboring hospitals, prompting an additional (indirect) rise in mortality rates there also. Spillovers due to individuals’ characteristics within the same treatment group are discussed in \citet{Ogburn:VanderWeele:2014}.
} First, we completely circumvent pitfalls regarding Type-1 interference, since our counterfactual scenario is generated without relying on untreated units. Second, we do not postulate the lack of interference among treated units and allow for any possible Type-2 interference. We avoid relying on the no-interference-among-treated assumption because of the likely presence of interference across both the temporal and cross-sectional dimensions in social science applications \citep{Xu:2023}. Consequently, our focus is on estimating the \textit{total} effect of the treatment, which encompasses both the direct effect on a unit from the treatment it receives and the indirect effects arising from the spillover and general equilibrium effects. We claim that in real-world scenarios with likely interaction between units, focusing on the total effect of the treatment is a sensible choice as it captures the actual impact on each unit under the realized treatment assignment mechanism. Under Assumption \ref{assumpt:sutva}, we can index the potential outcomes to the common treatment path: $\Y_{i,t}(1)$ indicates the potential outcome under the treatment, and $\Y_{i,t}(0)$ is the counterfactual outcome.   

The next assumption contributes to the definition of causal effects.

\begin{assumpt}[Additivity]
\label{assumpt:additivity}
We assume that the intervention produces an additive effect on the potential outcomes, i.e.,
\begin{small}
\begin{equation}
\label{eqn:additivity}
    \Y_{i,t}(1) = \Y_{i,t}(0) + \Delta_{i,t} .
\end{equation}
\end{small}
\end{assumpt}

This is not a restrictive assumption, since it holds after a suitable transformation of the data. For example, we can start from a multiplicative effect and then apply a logarithmic transformation to find an additive structure. Since inference is conducted using block-bootstrap, if it is necessary to express the effect in its original form, the inverse transformation can be applied to the bootstrap distribution, so as to recover the original multiplicative effect and its confidence interval. In conjunction with Assumption \ref{assumpt:model} (see below), this assumption implies that we effectively adopt a Partially Linear Model, akin to seminal causal machine learning methods \citep{chernozhukov2018double, Wager:Athey:2018}. The Partially Linear Model strikes a balance between structure and flexibility: the causal-effect component of the model remains simple and interpretable, while the untreated potential outcome can exhibit almost arbitrary complexity \citep{chernozhukov2024applied}.

Assumption \ref{assumpt:no_anticipation} rules out the possibility of anticipatory effects of the intervention on both the outcome and the covariates \citep{Abadie:Diamond:Hainmueller:2010, Borusyak:Jaravel:Spiess:2021}.

\begin{assumpt}[Absence of anticipation]   \label{assumpt:no_anticipation}
Let $\bb{X}_{i,t} = (\X_{i,t}^{(1)}, \dots, \X_{i,t}^{(m)})$ be an $m$-dimensional vector of covariates that are predictive of the outcome $i$ at time $t \leq t_0$. We assume: i) absence of anticipatory effects of the intervention on the covariates and the potential outcomes, i.e., $\Y_{i,t}(1)=\Y_{i,t}(0)$ and $\bb{X}_{i,t}(1)= \bb{X}_{i,t}(0)$ for all $t \leq t_0$; ii) future covariates do not affect current potential outcomes; iii) \textup{(optional)} some of the covariates remain unaffected by the policy in the post-intervention period (post-treatment exogeneity of the covariates), i.e., for all $t > t_0$, $\bb{X}_{i,t}(1)=\bb{X}_{i,t}(0)$.
\end{assumpt}

This assumption implies that in the pre-intervention period, the observed outcome corresponds to the potential outcome absent the policy, i.e., $\Y_{i,t} \equiv \Y_{i,t}(0)$ and that $\bb{X}_{i,t} \equiv \bb{X}_{i,t}(0)$ throughout all the analysis period. Additionally, Assumption \ref{assumpt:no_anticipation} precludes any anticipatory actions by agents, implying that they cannot alter or manipulate their outcomes prior to receiving the treatment. This assumption can be empirically tested by verifying whether pre-treatment effects are, on average, zero. Finally, post-treatment exogeneity of the covariates is not essential for identification, as in many applications, it is more realistic to use only lagged covariates. Thus, part (iii) of Assumption \ref{assumpt:no_anticipation} is relevant only in cases where post-treatment values of some covariates are needed to control for post-treatment confounders and these covariates can be considered as exogenous \citep{Brodersen:Gallusser:Koehler:Remy:Scott:2015}.\footnote{Although motivating covariates’ choice is often sufficient, it can be also verified by testing for the presence of treatment effects on each covariate: those that are significantly impacted by the intervention must be removed from the model.} 

The next identification assumption pertains to the potential outcomes model, 
similar to \citet{Carvalho:Masini:Medeiros:2018} and \citet{Masini:Medeiros:2021}, but instead of using control units, we rely on a dynamic specification based on lagged outcomes.

\begin{assumpt}[Dynamic potential outcomes model]
\label{assumpt:model}
    Denote with $\bb{X}_{i,t}^{(q)} = ( \bb{X}_{i,t}, \dots, \bb{X}_{i,t-q} )$ the collection of covariates up to time $t-q > 0$, and let  $\Y_{i,t-1}^{(p)}(0) = ( \Y_{i,t-1}(0), \dots, \Y_{i, t-1-p}(0) )$ be the collection of past potential outcomes up to time $t-p > 0$, with $p \in \{0, \dots, t-2 \}$ and $q \in \{0, \dots, t-1 \}$. We assume that, for all $t = 1, \dots, T$, the potential outcome absent the policy is as follows, \vspace{-5pt}
    \begin{small}
\begin{align}
    \label{eqn:model_assumpt}
    \Y_{i,t}(0) & = h \left( \Y^{(p)}_{i,t-1}(0), \bb{X}^{(q)}_{i,t} \right) + \epsilon_{i,t} 
\end{align}
\end{small}
where $h(\cdot)$ is some flexible function of the past lags of the outcome, the contemporaneous covariates, and the past lags of covariates; $\epsilon_{i,t}$ is a zero-mean, uncorrelated error term following a generic distribution $f(\cdot)$ 
\end{assumpt}


Notice that Equation (\ref{eqn:model_assumpt}) assumes poolability, i.e., homogeneous intercept and slope for all units, meaning that once we account for a highly predictive set of covariates and include temporal dynamics through lagged outcomes, the residual component is essentially random noise.\footnote{For panel data studies with small T and large N, it is usual to pool the observations \citep{baltagi2008econometric}.} We posit this assumption can be met in practice by using a subset of very predictive covariates out of a larger initial set selected \textit{ex ante} on the basis of domain knowledge. This approach ensures that major sources of heterogeneity among units are effectively addressed. In addition, we can provide empirical support for this assumption by examining the fit of the pooled model during the pre-intervention period. A good fit, indicated, for instance, by low Mean Squared Error (MSE), would suggest that the pooled model adequately captures the variations in the data. In cases where the researcher also wants to account for time-invariant, group-level unobserved heterogeneity (e.g., units’ fixed effects), the covariate set could be augmented by adopting recent approaches based on sufficient representations for categorical variables \citep{Johannemann:Hadad:Athey:Wager:2019}.

We remark that $h(\cdot)$ can be either a linear or a non-linear function. In the linear case, no additional assumptions are needed for the identification of causal effects. In the non-linear case, no further assumptions are required if the effect is defined only a single step ahead, as in our empirical application. Conversely, identification of non-linear multi-step-ahead causal effects is notoriously difficult and has not yet been fully explored in the literature (for time series in non-causal settings, see \citet{Chen:Yang:Hafner:2004}). Therefore, our paper also contributes to the formal identification of causal effects in this challenging context. 
Specifically, when $h(\cdot)$ is non-linear and the focus is on multi-step-ahead causal effects, we propose modeling the potential outcomes in the post-intervention period as follows.
\begin{assumpt} [Post-intervention non-linear multi-step-ahead model
]\label{assumpt:additional_models}
For a positive integer $k \geq 2$, denote by $\Y_{t_0+k|t_0}(0)$ the expected potential outcome absent the policy, given covariates and pre-intervention lags of the outcome, i.e., $\Y_{i,t_0+k|t_0}(0) = \E[\Y_{i,t_0+k}(0)|\Y_{i,t_0}^{(p)}, \bb{X}_{i,t_0+k}^{(q)}]$. The potential outcomes absent the policy at time $t_0+k$ are functions of their past lags up to $t_0$, their conditional expectations $\Y_{i,t_0+1|t_0}, \dots, \Y_{i,t_0+k-1|t_0}$, and the covariates $\bb{X}^{(Q)}_{i,t_0+k}$, with $Q = \max{ \{ k-1, q\} }$, 
\begin{small}
    \begin{align}        \label{eqn:direct_models}
        \Y_{i,t_0+k}(0) & = g_k \left(\Y_{i,t_0+k-1|t_0}(0), \dots, \Y_{i,t_0+1|t_0}(0), \Y_{i,t_0}^{(p)}, \bb{X}_{i,t_0+k}^{(Q)}\right) + \epsilon_{i,t_0+k}  
    \end{align}
\end{small}
where $\epsilon_{i,t_0+k}$ is a zero-mean, uncorrelated error term, and $g_k(\cdot)$ is a flexible function which can be different at each horizon $k$.
\end{assumpt}
Notice that Equation (\ref{eqn:direct_models}) does not impose any particularly restrictive assumption on the potential outcomes. Indeed, it merely formalizes the dependence structure between potential outcomes and their conditional expectations already implied by Equation (\ref{eqn:model_assumpt}), e.g., for $k = 2$ and $p = q = 0$, we have $\Y_{i,t_0+2}(0) = h(\Y_{i,t_0+1|t_0}(0) + \epsilon_{i,t_0+1}, \bb{X}_{i,t_0+2}) + \epsilon_{i,t_0+2} = g_2(\Y_{t_0+1|t_0}(0), \Y_{i,t_0}, \bb{X}_{i,t_0+2}, \bb{X}_{i,t_0+1}) + \epsilon_{i,t_0+2}$. 


The panel CV procedure---thoroughly described in Section \ref{sec:implementation}---which we propose to estimate the pre-intervention potential outcome model in Equation (\ref{eqn:model_assumpt}), will reveal whether the winner of the horse-race is a linear or non-linear model. If a linear model is selected, Equation (\ref{eqn:model_assumpt}) remains valid for all $t = 1, \dots, T$. Otherwise, we can use the non-linear specification in Equation (\ref{eqn:direct_models}) to identify and estimate post-intervention counterfactual outcomes after $t_0+1$.  

Finally, the potential outcome model defined by Equation (\ref{eqn:model_assumpt}) does not assume stationarity of the outcome variable. In similar contexts, studies often adopt a weaker assumption of conditional stationarity \citep{DHaultfoeuille:Hoderlein:Sasaki:2023, Hoderlein:White:2012}. Formally, let $f_{\Y_{i,t}(0)|\Y_{i,t-1}^{(p)}(0),\bb{X}_{i,t}^{(q)}}$ be the conditional density of the potential outcomes under control. Conditional stationarity assumes that this density remains unchanged one step ahead, i.e., $f_{\Y_{i,t}(0)|\Y_{i,t-1}^{(p)}(0),\bb{X}_{i,t}^{(q)}} = f_{\Y_{i,t+1}(0)|\Y_{i,t}^{(p)}(0),\bb{X}_{i,t+1}^{(q)}}$. In other words, its functional form, which matches the form of the error term, is time-invariant, although the conditional mean can still be time-varying.\footnote{Consider this simple (non-linear) example: $Y_t = \sin{Y_{t-1}^2} + \varepsilon_t$ where $\varepsilon_t \sim f(0,\sigma^2_{\epsilon})$. Under conditional stationarity, we have that $Y_t|Y_{t-1} \sim f(\sin{Y_{t-1}^2}, \sigma^2_{\epsilon})$, $Y_{t+1}|Y_{t} \sim f(\sin{Y_{t}^2}, \sigma^2_{\epsilon})$ and so on. Notice that while the conditional mean is time-varying, the form of the distribution $f(\cdot)$ remains the same. More generally, in our setting of interest, a short panel with small T and large N, non-stationarity is not an issue of particular concern \citep{baltagi2008econometric}. Nevertheless, as we will show below, the simulation results reported in Supplemental Appendix \ref{app:simulation} demonstrate that even in the case of an explosive autoregressive coefficient (i.e., when the coefficient of the past lag of the outcome $Y_{t-1}$ is set at $\phi = 1.2$), both the bias and the coverage rates of the MLCM are in line with the results obtained for stationary processes (i.e., when $\phi = 0.8$).} This means that the error term would be distributed in the same way during the post-intervention period. In our case, as we explain in Section \ref{subsec:estimands}, we target the causal effect at the individual level and specifically its expected value conditional on past information. Therefore, for identification, estimation, and inference, we do not require the error distribution to remain unchanged after the intervention. Instead, it is sufficient that the error term remains uncorrelated and has a mean of zero, a condition already embedded in Assumptions \ref{assumpt:model} and \ref{assumpt:additional_models}. In other words, we only require that, in the absence of the policy, the potential outcomes model would continue to be correctly specified.  
This implies the absence of other unforecastable shocks or policies affecting the outcome of interest in the post-intervention window.
While it is not possible to explicitly test for this, we can assess how well the model fits the pre-intervention data via the rigorous panel CV procedure (see Section \ref{sec:implementation}).

\subsection{Causal estimands}
\label{subsec:estimands}

We introduce novel estimands that could be of general interest for observational studies based on panel data in the absence of control units. The causal estimands discussed here pertain to Case (i) defined above, namely, when the treatment affects all, or most, of the available units, simultaneously. In Supplemental Appendix \ref{app:additional_estimands}, we also define causal estimands for the Case (ii) scenario with violations of the no-interference assumption.

We begin with the unconditional individual causal effect estimand.
\begin{defi}
    \label{def:uce}
    For any $k\geq 1$, the unconditional causal effect of the policy on unit $i$ at time $t_0+k$ is,
    \begin{small}
    \begin{equation}
        \label{eqn:uce}
        \Delta_{i,t_0+k} = \Y_{i,t_0+k}(1) - \Y_{i,t_0+k}(0).
    \end{equation}
    \end{small}
\end{defi}
Note that this estimand arises naturally as a consequence of Assumption (\ref{assumpt:additivity}). However, identification, estimation, and inference of this causal effect would require stringent assumptions (e.g., strong stationarity of the potential outcome distribution). Therefore, our focus is on its conditional mean. 
\begin{defi}
    \label{def:ind_effect}
    For any $k\geq 1$ and some $p \in \{0, \dots, t_0-1 \}$, $q \in \{ 0, \dots, t_0+k-1 \}$ with $Q = \max{ \{k-1,q \} }$, 
    the expected individual effect of the policy, conditional on past outcomes and covariates, at time $t_0+k$ is,
    \begin{small}
    \begin{align}        \label{eqn:ind_effect}
        \nonumber
        \tau_{i, t_0+k} & = \E[\left( \Y_{i,t_0+k}(1) - \Y_{i,t_0+k}(0) \right) | \Y_{i,t_0}^{(p)}, \bb{X}_{i,t_0+k}^{(Q)}] \\
        \nonumber
        & = \Y_{i,t_0+k}(1) - \E[\Y_{i,t_0+k}(0) | \Y_{i,t_0}^{(p)}, \bb{X}_{i,t_0+k}^{(Q)}] \\
        & = \Y_{i,t_0+k}(1) - \Y_{i,t_0+k|t_0}(0).
    \end{align}
    \end{small}
\end{defi}
This estimand emphasizes that, because the potential outcomes are temporally related, 
the causal effect of the policy should be defined conditionally on past information. Furthermore, in many situations, the full distribution of $\Delta_{i,t_0+k}$ is not of interest unless we specifically target causal effects at the median or other quantiles. Typically, the mean effect is the primary concern, making $\tau_{i,t_0+k}$ a more suitable summary measure of policy effects.   
The following theorem establishes the identification conditions for the individual effect $\tau_{i,t_0+k}$. 
\begin{theo}
For $k = 1$, $\tau_{i,t_0+k}$ is identified under Assumptions \ref{assumpt:no_anticipation} and \ref{assumpt:model}. Under a linear potential outcomes model, the identifying conditions for $k\geq 2$, are the same as for $k = 1$. If the potential outcomes model is non-linear, $\tau_{i,t_0+k}$ is identified under Assumptions \ref{assumpt:no_anticipation} and \ref{assumpt:additional_models}.
\end{theo}

\begin{proof}
    We illustrate the proof in a simple case where the potential outcome absent the policy depends only on one previous lag and contemporaneous covariates, i.e., Equations (\ref{eqn:model_assumpt}) and (\ref{eqn:direct_models}) become, respectively,
$$\Y_{i,t}(0) = h(\Y_{i,t-1}(0), \bb{X}_{i,t}) + \epsilon_{i,t}$$
$$\Y_{i,t_0+k}(0) = g_k(\Y_{i,t_0+k-1|t_0}(0), \dots, \Y_{i,t_0+1|t_0}(0), \Y_{i,t_0}, \bb{X}_{i,t_0+k}) + \epsilon_{i,t}$$
corresponding to the case $p = q = 0$. The general proof is reported in Supplemental Appendix \ref{app:proof_general}. 
The proof proceeds by induction and is organized distinguishing the two cases $k = 1$ and $k = 2$. 

\vspace{5pt}

\noindent

\begin{itemize}
    \item $k = 1$. In the expression $\tau_{i,t_0+1} = \Y_{i,t_0+1}(1) - \Y_{i,t_0+1|t_0}(0)$, the first term $\Y_{i,t_0+1}(1)$ is the observed outcome under the policy and thus is immediately identified. Since we are in the case $p=q=0$, we have that $Q = \max{ \{k-1, q \} } = 0$, so the second term can be written as 
    \begin{small}
    \begin{align*} 
        \Y_{i,t_0+1|t_0}(0) & = \E[\Y_{i,t_0+1}(0)|\Y_{i,t_0}, \bb{X}_{i,t_0+1}] & \text{Def. of $\Y_{i,t_0+1|t_0}(0)$ for $p = Q = 0$} \\ 
        & = \E[h(\Y_{i,t_0}(0), \bb{X}_{i,t_0+1}) |\Y_{i,t_0}, \bb{X}_{i,t_0+1}]  & \text{Assumption \ref{assumpt:model} - Eq.(\ref{eqn:model_assumpt})} \\       
        & = h(y_{i,t_0}(0), \bb{x}_{i,t_0+1}) = y_{i,t_0+1|t_0}(0).
    \end{align*}
    \end{small}
    The last expression follows from the fact that $h(\cdot)$ is deterministic once we observe $\Y_{i,t_0}(0) = y_{i,t_0}(0)$ and $\bb{X}_{i,t_0+1} = \bb{x}_{i,t_0+1}$. As a result, $\Y_{i,t_0+1|t_0}(0)$ is identified from observed data, and so is $\tau_{i,t_0+1}$. Notice that in this proof, we never used the linearity of the potential outcome model. Thus, the proof at $k = 1$ is the same even when $h(\cdot)$ is non-linear.  
   
    \item $k = 2$. In the expression $\tau_{i,t_0+2} = \Y_{i,t_0+2}(1) - \Y_{i,t_0+2|t_0}(0)$, the first term $\Y_{i,t_0+2}(1)$ is the observed outcome under the policy. Thus, the identification of the causal effect depends solely on the second term, $\Y_{i,t_0+2|t_0}(0)$. We now distinguish between a linear and a non-linear model specification:
    \end{itemize}
    \begin{itemize}    
        \item[-] Linear case. Assume the following linear model $h(\Y_{i,t-1}(0), \bb{X_t}) = b_1\Y_{i,t-1}(0) + \bb{b_2} \bb{X}_{i,t} $, where $\bb{b_2}$ is an $m-$dimensional vector of covariates' coefficients. Since we are in the case $p=q=0$, we have that $Q = \max{ \{1, 0 \} } = 1$, so the term $\Y_{i,t_0+2|t_0}(0)$ can be written as,
        \begin{small}
        \begin{align*}
         \Y_{i,t_0+2|t_0}(0)  & = \E[\Y_{i,t_0+2}(0)|\Y_{i,t_0}, \bb{X}_{i,t_0+2}, \bb{X}_{i,t_0+1}] & \text{Def. $\Y_{i,t_0+2|t_0}(0)$ for $p = 0, Q =1$}  \\         
        & = \E[h(\Y_{i,t_0+1}(0), \bb{X}_{i,t_0+2})|\Y_{i,t_0}, \bb{X}_{i,t_0+2}, \bb{X}_{i,t_0+1}] & \text{Assumption \ref{assumpt:model} - Eq.(\ref{eqn:model_assumpt})} \\
        & = \E [b_1 \Y_{i,t_0+1}(0) + \bb{b}_2 \bb{X}_{i,t_0+2}|\Y_{i,t_0}, \bb{X}_{i,t_0+2}, \bb{X}_{i,t_0+1}] & \text{Linear model} \\
        & = b_1 \E[\Y_{i,t_0+1}(0)|\Y_{i,t_0}, \bb{X}_{i,t_0+1}] + \bb{b}_2 \bb{x}_{i,t_0+2} & \text{Assumption \ref{assumpt:no_anticipation}} \\
        & = b_1 y_{i,t_0+1|t_0}(0) + \bb{b}_2 \bb{x}_{i,t_0+2}.      
    \end{align*}
    \end{small}
    \item[-] Non-linear case. The identification of $\Y_{i,t_0+2|t_0}(0)$ can be based on the formulation for the post-intervention potential outcome model defined by Equation (\ref{eqn:direct_models}),
    \begin{small}    
    \begin{align*}
         \Y_{i,t_0+2|t_0}(0)  & = \E[\Y_{i,t_0+2}(0)|\Y_{i,t_0}(0), \bb{X}_{i,t_0+2}, \bb{X}_{i,t_0+1}] \hspace{70pt} \text{Def. $\Y_{i,t_0+2|t_0}(0)$ for $p = 0, Q =1$} \\         
        & = \E[g_2(\Y_{i,t_0+1|t_0}(0),\Y_{i,t_0}, \bb{X}_{i,t_0+2})|\Y_{i,t_0}, \bb{X}_{i,t_0+2}, \bb{X}_{i,t_0+1}] \hspace{50pt} \text{Assumption \ref{assumpt:additional_models} - Eq.(\ref{eqn:direct_models})}  \\
        & = \E[g_2(h(\Y_{i,t_0}, \bb{X}_{i,t_0+1}),\Y_{i,t_0}, \bb{X}_{i,t_0+2})|\Y_{i,t_0}, \bb{X}_{i,t_0+2}, \bb{X}_{i,t_0+1}]  \\
        & =g_2(y_{i,t_0+1|t_0}(0), y_{i,t_0},\bb{x}_{i,t_0+2}).
    \end{align*}
    \end{small}
    The last expression follows from the fact that $g_2(\cdot)$ is deterministic once we observe $\Y_{i,t_0} = y_{i,t_0}$, $\bb{X}_{i,t_0+1} = \bb{x}_{i,t_0+1}$ and  $\bb{X}_{i,t_0+2} = \bb{x}_{i,t_0+2}$. In addition, even though Assumption \ref{assumpt:no_anticipation} was not explicitly mentioned in this part of the proof, notice that it is implicit in Assumption \ref{assumpt:additional_models}, as by Equation (\ref{eqn:direct_models}), the potential outcomes never depend on future covariates.
    \end{itemize}
     
The proof for a generic time $t_0+k$ follows analogously by induction. 
\end{proof}

The following estimands are unit-averages of the individual effects defined above. For the reasons previously stated, we focus on averaging the conditional individual effects, which can be considered the panel analog of the ATE. We also frame them from a finite-sample perspective, as the MLCM is designed for statistical and econometric settings where the treatment affects, either directly or indirectly, the entire population under study.\footnote{For further details on the difference between the finite-sample and superpopulation perspectives in causal inference, see \citet{Imbens:Rubin:2015}. In observational panel settings, only \citet{Rambachan:Roth:2023} defined causal estimands when the entire population is observed, but their approach relies on control units.
}

\begin{defi}
\label{def:ate}
For any $k\geq 1$, the ATE at time $t_0+k$ is defined as the average of the individual effects across all $i$ units in the panel, with $i = 1, \dots, N$,
\begin{equation}
\label{eqn:ate}
     \tau_{t_0+k} = \frac{1}{N} \sum_{i = 1}^N \tau_{i, t_0+k} = \frac{1}{N} \sum_{i = 1}^N \Y_{i,t_0+k}(1) - \Y_{i,t_0+k|t_0}(0) .
\end{equation}                                     
\end{defi}
Note that in this benchmark scenario (Case (i) where only treated units are available), the term $\tau_{t_0+k}$ corresponds to the Average Treatment effect on the Treated (ATT).

The subsequent estimand measures the average individual effect within a subset of units with the same values of selected covariates. This is commonly referred to as the CATE, and it is of particular interest when there is reason to believe that the intervention has produced heterogeneous effects on different subpopulations of units as defined by their distinct characteristics. Following existing literature on CATE in high-dimensional settings with a mix of discrete and continuous covariates \citep{Fan:Hsu:Lieli:Zhang:2022, Knaus:Lechner:Strittmatter:2021, Chernozhukov:Demirer:Duflo:Fernandez:2018}, we focus on its low-dimensional summary called “group-average treatment effect”. 

\begin{defi}
\label{def:cate}
Let $G_{i,t}$ denote a set of individual characteristics and indicate with $N_g$ the number of units in the population having $G_{i,t} = g$. For any $k \geq 1$, the CATE at time  $t_0+k$ is defined as,
\begin{equation}
    \label{eqn:cate}
    \tau_{t_0 + k}(g) = \frac{1}{N_g} \sum_{i: G_{i,t_0+k} = g} \tau_{i, t_0 +k} .
\end{equation}
\end{defi}

Note that the set of candidate conditioning variables $G_{i,t}$ used to determine CATEs does not necessarily have to match those utilized for counterfactual forecasting. This is because the variables that predict outcomes can, and often do, differ at least partially from those that predict treatment effect heterogeneity. We also remark that in a general panel setting with more than one post-treatment periods, the above definitions imply the existence of vectors representing estimated average effects, i.e., $ (\tau_{t_0+1}, \dots, \tau_T )$ and $( \tau_{t_0+1}(g), \dots, \tau_T (g))$. Furthermore, researchers are sometimes also interested in temporal aggregations of such effects. 

\begin{defi}
\label{def:ate_avg}
The temporal average ATE and CATE are defined, respectively, as,
\begin{align*}    
    \bar{\tau} = \frac{1}{T - t_0} \sum_{k = 1}^{T - t_0} \tau_{t_0+k} & &  \bar{\tau}(g) = \frac{1}{T - t_0} \sum_{k = 1}^{T - t_0} \tau_{t_0+k}(g) 
\end{align*}
\end{defi}
 
We now introduce the following Theorem.
\begin{theo}
For $k = 1$, ATE and CATE are identified under Assumptions \ref{assumpt:sutva}, \ref{assumpt:no_anticipation}, and \ref{assumpt:model}. Under a linear potential outcomes model, the identifying conditions for $k \geq 2$, are the same as for $k = 1$; if the potential outcomes model is non-linear, ATE and CATE are identified under Assumptions \ref{assumpt:sutva}, \ref{assumpt:no_anticipation}, and \ref{assumpt:additional_models}.
\end{theo}
\begin{proof}
    The proof follows directly from the previous one. By Definitions \ref{def:ate} and \ref{def:cate}, we have that both $\tau_{t_0+k}$ and $\tau_{t_0+k}(g)$ are aggregations of $\tau_{i,t_0+k}$ across different sets of units (all the $N$ units in the panel for ATE, subgroups sharing the same characteristics for CATE). Thus, their unit-average is also identified under the same set of assumptions. We also add Assumption \ref{assumpt:sutva} because in this case we are considering aggregation of units, so the effects must be comparable (if Assumption \ref{assumpt:sutva} is violated, we would estimate the effects of different forms of treatment that cannot be aggregated). This logic extends to the temporal average effects in Definition \ref{def:ate_avg}.
\end{proof}

\subsection{Estimators}
\label{subsec:estimators}
We now introduce estimators of the causal quantities defined in Section \ref{subsec:estimands}.  Note that imputing future counterfactual outcomes based on past information and covariates is essentially a forecasting problem. Therefore, in the following definitions, the estimator $\widehat{\Y}_{t_0+k|t_0}$ represents the $k$-step-ahead forecast based on the model trained in the pre-intervention period. We first define an estimator for the individual effect at time $t_0+1$.
\begin{defi}
\label{def:hat_ind}
For some $p = 0, \dots, t_0-1$ and $q = 0, \dots, t_0$ with $Q = \max{ \{ 0,q\} } $, an estimator of $\tau_{i,t_0+1}$ is the difference between the observed outcome under the policy and the estimated $1$-step-ahead forecast, conditional on past information and contemporaneous covariates,
\begin{small}
\begin{align}
    \label{eqn:hat_mcite_1}
    \nonumber    \widehat{\tau}_{i,t_0+1} & = \Y_{i,t_0+1}(1) - \widehat{\Y}_{i,t_0+1|t_0}(0) \\
    & = \Y_{i,t_0+1}(1) - \E \left[\widehat{h}(\Y_{i,t_0}^{(p)}, \bb{X}_{i,t_0+1}^{(q)}) \big|\Y_{i,t_0}^{(p)}, \bb{X}_{i,t_0+1}^{(Q)} \right] \\
    & = \Y_{i,t_0+1}(1) - \widehat{h}(y_{i,t_0}^{(p)}(0), \bb{x}^{(q)}_{i,t_0+1})
\end{align}
\end{small}
where $\widehat{h}(\cdot)$ denotes that the $1$-step-ahead forecast is based on the optimized ML parameters.
\end{defi}
A detailed description of the estimation algorithm for $h(\cdot)$ will be given in Section \ref{sec:implementation}. Although our empirical application only evaluates causal effects at time $t_0+1$, one might also be interested in estimating effects over multiple post-intervention periods, as in our simulations and replication study. There are two main strategies for generating multi-step-ahead forecasts \citep{Chevillon:2007, Forecasting:2022}: the \textit{recursive} approach, where each forecast is defined and estimated using previous forecasts, and the \text{direct} approach, where forecasts are produced by estimating separate models for each forecast horizon. In the case of linear models, the recursive approach yields more efficient parameter estimates, especially when the model is correctly specified and for a long forecasting horizon \citep{Pesaran:Pick:Timmermann:2011}. However, with non-linear models, the recursive approach is known to be asymptotically biased, making the direct approach often preferable \citep{Bontempi:Taieb:2011}. One criticism of the direct forecasting approach is that it assumes independence between the forecasts, as each one is based on a separate model. \textit{Hybrid} forecasts, instead, represent a combination of the recursive and direct approaches, effectively integrating the strengths of both strategies \citep{Forecasting:2022}. Like the direct approach, hybrid strategies require estimating separate models for each forecasting horizon. However, similar to the recursive approach, each model also includes the direct forecast from the previous step. This method offers full flexibility by allowing different models for each time horizon, while still accounting for the dependency between direct forecasts. This is exactly what Equation (\ref{eqn:direct_models}) captures: each $\Y_{i,t_0+k-1|t_0}(0), \dots, \Y_{i,t_0+1|t_0}(0)$ is a direct forecast (i.e., the expected potential outcome up to $k-1$ steps ahead from $t_0$), and $\Y_{i,t_0+k}(0)$ is then modeled as a flexible function of these direct forecasts, along with observed past lags and covariates. 

Therefore, depending on whether the potential outcome model is linear or non-linear, the next definition outlines two different estimators for multi-step-ahead evaluations.
\begin{defi}
For $k \geq 2$ and for some $p = 0, \dots, t_0-1$ and $q = 0, \dots, t_0+k-1$ with $Q = \max{ \{ k-1,q\} } $, an estimator for $\tau_{i,t_0+k}$ when $h(\cdot)$ is linear is given by,
\begin{small}
\begin{align}    \label{eqn:hat_mcite_k_lin}    \widehat{\tau}^{lin}_{i,t_0+k} & = 
    \Y_{i,t_0+k}(1) - \widehat{\Y}_{i,t_0+k|t_0}(0) \\ 
    & = \Y_{i,t_0+k}(1) - \E \left[\widehat{h}(\Y_{i,t_0+k-1}^{(p)}(0), \bb{X}_{i,t_0+k}^{(q)}) \big|\Y_{i,t_0}^{(p)}, \bb{X}_{i,t_0+k}^{(Q)} \right]     
\end{align}
\end{small}
whereas, if $h(\cdot)$ is non-linear an estimator for $\tau_{i,t_0+k}$ is given by,
\begin{small}
\begin{align}    \label{eqn:hat_mcite_k_nonlin}    \widehat{\tau}^{nl}_{i,t_0+k} & = 
    \Y_{i,t_0+k}(1) - \E \left[\widehat{g}_k(\Y_{i,t_0+k-1|t_0}(0), \dots, \Y_{i,t_0+1|t_0}(0), \Y_{i,t_0}^{(p)}, \bb{X}_{i,t_0+k}^{(q)}) \big|\Y_{i,t_0}^{(p)}, \bb{X}_{i,t_0+k}^{(Q)} \right]
\end{align}
\end{small}
where $\widehat{h}(\cdot)$ and $\widehat{g}_k(\cdot)$ denote that the $k$-step-ahead forecasts are based on the optimized ML parameters.
\end{defi}
 A detailed description of the estimation algorithm for $g_k(\cdot)$ will be given in Section \ref{sec:implementation}. To clarify how multi-step-ahead forecasts are computed and to highlight the differences between the linear and non-linear case, we provide the following example.
\begin{example}
    Consider the simple case $p = q = 0$, and let $\widehat{h}(\Y_{i,t_0+k-1}(0), \bb{X}_{i,t_0+k}) = \widehat{b}_1 \Y_{i,t_0+k-1}(0) + \widehat{\bb{b}}_2 \bb{X}_{i,t_0+k}$. For $k = 2$, the $2$-step-ahead forecast $\widehat{\Y}_{i,t_0+2|t_0}(0)$ is given by
    \begin{small}
    \begin{align*}      \widehat{\Y}_{i,t_0+2|t_0}(0) & = \E[\widehat{\Y}_{i,t_0+2}(0)|\Y_{i,t_0}, \bb{X}_{i,t_0+2}, \bb{X}_{i,t_0+1} ] \\
      & = \E[\widehat{h}(\widehat{\Y}_{i,t_0+1}(0), \bb{X}_{i,t_0+2})|\Y_{i,t_0}, \bb{X}_{i,t_0+2}, \bb{X}_{i,t_0+1} ] \\
      & = \widehat{b}_1 \E[\widehat{\Y}_{i,t_0+1}(0)|\Y_{i,t_0}, \bb{X}_{i,t_0+1}] + \widehat{\bb{b}}_2 x_{i,t_0+2} \\
      & = \widehat{b}_1 \widehat{\Y}_{i,t_0+1|t_0} + \widehat{\bb{b}}_2 x_{i,t_0+2} = \widehat{h}(\widehat{\Y}_{i,t_0+1|t_0}, \bb{x}_{i,t_0+2})
    \end{align*}
    \end{small}
    Multi-step-ahead forecasts in the linear case are then computed by recursive substitution, plugging-in the previous forecast (and updated covariates) in the already estimated model $\widehat{h}(\cdot)$.  
    Now, assume that the panel CV procedure selects a non-linear model for the pre-intervention period. The $1$-step-ahead forecast $\widehat{\Y}_{i,t_0+1|t_0}(0)$ can be performed easily by following Equation (\ref{eqn:hat_mcite_1}). At $k = 2$, we then re-estimate a model $\widehat{g}_2(\widehat{\Y}_{i,t_0+1|t_0}(0), \Y_{i,t_0}, \bb{X}_{i,t_0+2}, \bb{X}_{i,t_0+1})$ that includes the forecasted counterfactual outcome, as in Equation (\ref{eqn:direct_models}). Thus, the $2$-step-ahead forecast $\widehat{\Y}_{i,t_0+2|t_0}(0)$ is given by,
    \begin{small}
    \begin{align*}      \widehat{\Y}_{i,t_0+2|t_0}(0) & = \E[\widehat{\Y}_{i,t_0+2}(0)|\Y_{i,t_0}, \bb{X}_{i,t_0+2}, \bb{X}_{i,t_0+1} ] \\
    & = \E[\widehat{g}_2(\widehat{\Y}_{i,t_0+1|t_0}, \Y_{i,t_0}, \bb{X}_{i,t_0+2}, \bb{X}_{i,t_0+1})|\Y_{i,t_0}, \bb{X}_{i,t_0+2}, \bb{X}_{i,t_0+1}] \\
    & = \widehat{g}_2(\widehat{h}(y_{i,t_0}, \bb{x}_{i,t_0+1}), y_{i,t_0}, \bb{x}_{i,t_0+2}).
    \end{align*}
    \end{small}
\end{example}    
Building on Equations (\ref{eqn:hat_mcite_1}-\ref{eqn:hat_mcite_k_nonlin}), the next definition summarizes the ATE and CATE estimators under the MLCM.
\begin{defi}
\label{def:hat_ate}
For any positive integer $k$, finite-sample estimators for the ATE and CATE at time $t_0 + k$  (\ref{eqn:ate}) are, respectively,
\begin{small}
\begin{align}
\label{eqn:hat_ate_cate}  \widehat{\tau}_{t_0+k} = \frac{1}{N} \sum_{i = 1}^N \widehat{\tau}_{i, t_0+k} & &\widehat{\tau}_{t_0 + k}(g) = \frac{1}{N_g} \sum_{i: G_{i,t_0+k} = g} \widehat{\tau}_{i, t_0 +k}.
\end{align}
\end{small}
Finally, the estimators for the temporal average ATE and CATE are, respectively,
\begin{small}
\begin{align}    \label{eqn:hat_ate_avg} \widehat{\bar{\tau}} = \frac{1}{T - t_0} \sum_{k = 1}^{T - t_0} \widehat{\tau}_{t_0+k} & & \widehat{\bar{\tau}}(g) = \frac{1}{T - t_0} \sum_{k = 1}^{T - t_0} \widehat{\tau}_{t_0+k}(g).
\end{align}
\end{small}
\end{defi}

Inference on the estimated causal effects defined above is conducted using block-bootstrap. Refer to Supplemental Appendix \ref{app:bootstrap} for a detailed description of the block-bootstrap algorithms used to derive confidence intervals for the ATE and CATEs. Finally, we stress that deriving the theoretical properties of the above-defined estimators would deviate from the intended nature of the methodology, since the MLCM can be implemented with any supervised ML algorithm, and it is unknown in advance which one will outperform the others.\footnote{To use Breiman's words: ``Nowhere is it written on a stone tablet what kind of model should be used to solve problems involving data." \citep{breiman2001statistical}.} Instead, in Section \ref{subsec:simulation}, we perform a simulation study showing that the MLCM can achieve forecast unbiasedness and is able to detect causal effects even in challenging econometric environments with non-linearities and irrelevant covariates included in the model specification. 


\section{The Machine Learning Control Method}
\label{sec:implementation}

\subsection{Departures from the standard machine learning approach}
\label{subsec:departures}
Supervised ML techniques primarily aim to minimize the out-of-sample prediction error, generalizing well on unseen data. The degree of flexibility is the result of a trade-off: increased flexibility can enhance in-sample fit but may diminish out-of-sample fit due to overfitting. ML algorithms tackle this trade-off by relying on empirical tuning to choose the optimal level of complexity. 

The standard ML approach is to randomly split the sample into two sets, containing, for instance, 2/3 and 1/3 of observations. One then uses the first set to train ML algorithms (training set) and the second to test them (testing set). This introduces a ``firewall'' principle: none of the data involved in generating the prediction function is used to evaluate it \citep{Mullainathan:Spiess:2017}. The out-of-sample performance of the model on the unseen (held-out) data of the testing set can be considered a reliable measure of the ``true'' performance on future data. In order to solve the bias-variance trade-off and prevent overfitting, one can rely on automatic tuning using tools such as random k-fold CV on the training sample to select the best-performing values of the tuning parameters in terms of an \textit{a priori} defined metric, such as the MSE. 

We depart from this standard ML routine and reorient it towards the counterfactual forecasting goal. First, we do not randomly split the data, but we train, tune, and evaluate the models only on the pre-treatment data (Design Stage); then, we use the final selected model to forecast counterfactual post-treatment outcomes. In this forecasting perspective, the unseen data on which the ML models must generalize well are not the outcomes of different units (as in typical out-of-sample prediction tasks), but future observations of the outcome for the same set of units employed to train the models. Stated differently, the aim is to make ML models learn as best as possible the pre-intervention outcome trend for each treated unit, so as to predict the best possible counterfactual outcome under the no-treatment scenario. The key implication of this unconventional ML setup is the shift in focus: the primary concern becomes ensuring unbiasedness in forecasts, rather than focusing only on forecast accuracy.

Second, and related, we do not carry out hyperparameter tuning and model selection with random k-fold CV. The panel dimension creates an additional challenge regarding how to implement CV, because standard CV does not account for the temporal structure of the data \citep{arkhangelsky2024causal}. To address this challenge, we introduce a resampling technique suited for forecasting tasks on panel data—panel CV—which is described below.

Finally, a key concern in ML regards the trade-off between accuracy and interpretability. Such a trade-off is relevant when ML is used for tasks that take into consideration transparency aspects. In principle, our method can be used with any supervised ML routine, including black-box techniques like deep neural networks. However, maintaining some degree of transparency can be important, because higher interpretability of the estimated counterfactuals bolsters the credibility of the proposed approach \citep{Abadie:2021}. In practice, users should decide on the basis of a comparative assessment across a mix of models characterized by different layers of complexity by balancing any improvements in performance from complex models against the loss of interpretability associated with their use.

\subsection{Implementation}
\label{subsec:implementation}
The MLCM is rooted in the idea of developing a ML forecasting model that can closely reproduce the outcome trajectories of treated units in the pre-intervention window, so that any post-intervention divergence between the observed and forecasted outcomes can be attributed to the treatment under the identification assumptions. To achieve this, the implementation of the MLCM requires ten empirical steps, which are divided into the Design Stage and the Analysis Stage. The full process is summarized in Box \ref{box:implementation} and described in detail below.

In the Design Stage, the first step involves the selection of supervised ML algorithms that will play the horse-race of performance testing on the pre-treatment data. This competition among multiple methods is conceptually analogous to the way the best ML learner is selected in the double/debiased ML method \citep{chernozhukov2018double}.\footnote{The MLCM can leverage any supervised ML algorithm, including deep learning techniques, some of which, such as RNNs and transformers, are explicitly designed to learn temporal dynamics; however, they cannot be employed in the panel settings we focus on (small T, large N), as they require a large number of time periods to be applicable. It is also possible to stack several different ML algorithms and form complex ensemble learners, but this would come at the cost of a substantial loss in transparency.} 

In the second step, we recommend deploying some tweaks involving feature pre-selection and engineering drawing from consolidated practices in applied predictive modeling \citep{Kuhn:Johnson:2013}. A crucial aspect of this pre-processing is the strategy for selecting predictors. There are two main schools of thought: those who advocate for purely data-driven selection argue that one should build a dataset as large as possible, and then let the algorithm autonomously decide which variables matter for the prediction task. Others stress the importance of subject matter knowledge: the researcher should select \textit{ex ante} the relevant predictors, and then feed only those to the algorithm. The rationale is that subject matter knowledge can separate meaningful from irrelevant information, eliminating detrimental noise and enhancing the underlying signal \citep{Kuhn:Johnson:2013}. 
We propose a hybrid approach: build a large initial dataset on the basis of domain knowledge, then adopt preliminary and data-driven variable selection criteria to drop non-informative predictors.

As outlined in the causal framework, counterfactual forecasting is carried out by using an information set mainly comprising lagged values of outcomes and covariates. However, determining the optimal number of lags to include is an empirical question. We recommend including at least two lagged values of both outcomes and covariates, and then using a data-driven approach to select a subsample of the most relevant features. The latter step allows for a reduction of the risk of overfitting and degradation of performances as well as facilitating interpretability. Finally, feature engineering and data pre-processing matter too because how the predictors enter into the model is also important \citep{Kuhn:Johnson:2013}.

To carry out model selection and validation on the pre-treatment sample, we propose a panel CV approach. Using an alternative CV procedure is necessary since ML methods do not natively handle longitudinal data and are designed for predicting rather than forecasting. 

\begin{figure}[h!]
    \centering
    \caption{Panel cross-validation (one-step-ahead forecasting). \textit{Notes}: this procedure is carried out using exclusively pre-treatment data. Light gray observations form the training sets; dark gray ones constitute the test sets. All the training windows also include past outcomes in the set of predictors.}
    \label{fig:pcv}    
    \includegraphics[scale = 0.17]{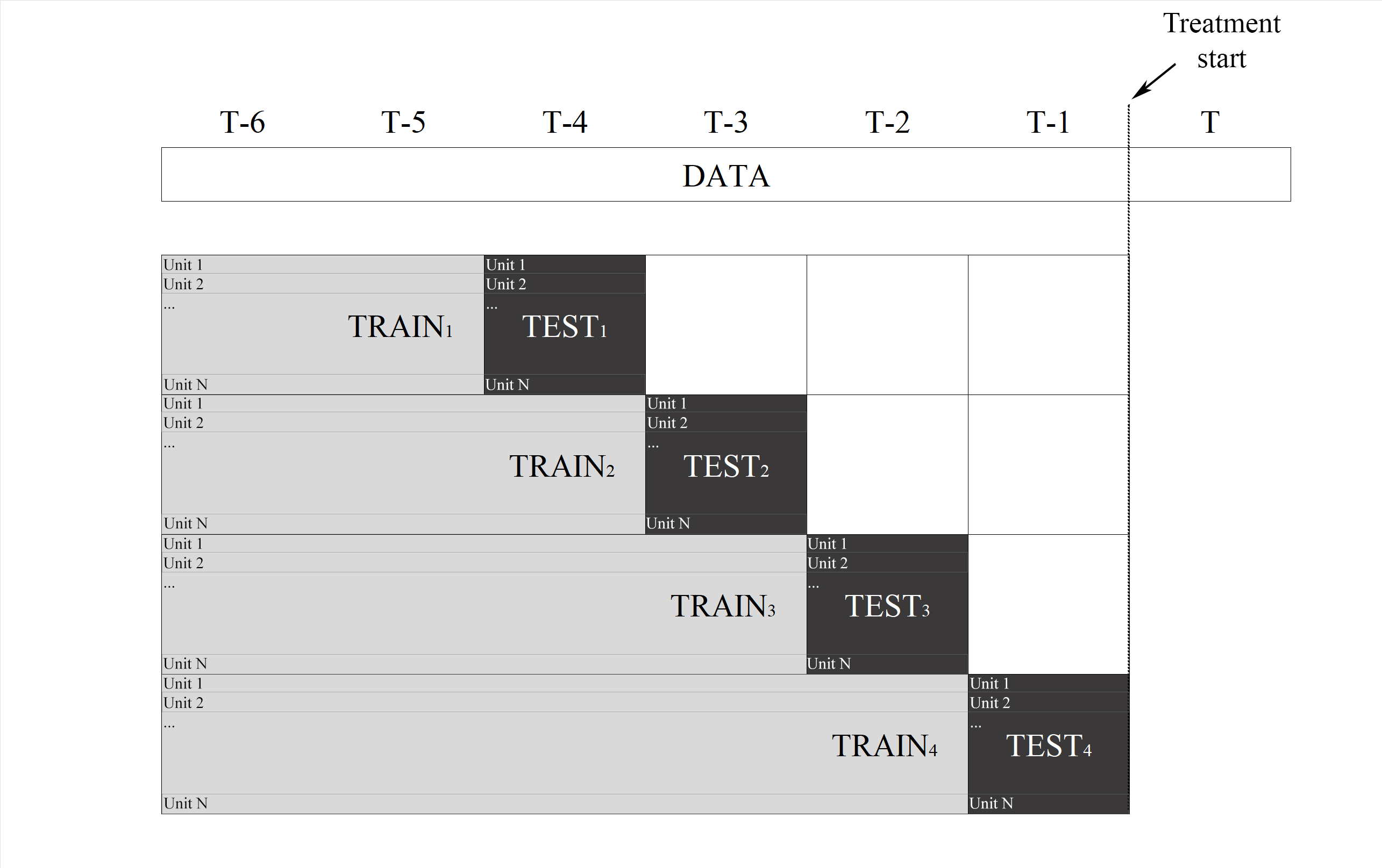}
\end{figure}

\vspace{10pt}
\begin{footnotesize}
\begin{spacing}{1}
    \begin{MyBox}{\hspace{5.cm}{Box \ref{box:implementation}: MLCM implementation}}{1} 
\label{box:implementation}
\begin{center}
    \textbf{Preliminary} 
    \vspace{2mm} 
    \hrule
\end{center} 
\vspace{1mm}
\begin{enumerate}[start=0]
    \item \textbf{Data splitting.} Split the full sample based on the treatment date: employ only pre-treatment data throughout the Design Stage; use post-treatment data only for estimating causal effects in the Analysis Stage.
\end{enumerate} 
\hrule
\begin{center}
    \textbf{Design Stage} 
    \vspace{2mm}
    \hrule
\end{center} 
\begin{enumerate}
    \item \textbf{Algorithm selection.} Select one or more supervised ML algorithms.
    \item \textbf{Principled input selection.} Build a large initial dataset on the basis of domain knowledge. To maximize forecasting performances, it is possible to use feature engineering, feature selection, and heuristic rules to pre-process the data and select a subsample of the most relevant predictors.
    \item \textbf{Panel cross-validation.} For each selected algorithm, tune hyperparameters via panel CV (see Figure \ref{fig:pcv} and Algorithms \ref{algorithm_PCV} and \ref{algorithm_multi_forecast}).
    \item \textbf{Performance assessment.} Assess average performance metrics (e.g., MSE) for all the selected algorithms and check what is the best-performing version of the MLCM.
    \item \textbf{Diagnostic and placebo tests.} Implement a battery of diagnostic and placebo tests to bolster the credibility of the research design.    
\vspace{2mm} 
\hrule
\begin{center}
\textbf{Analysis Stage} \hspace{0.775cm}
\vspace{2mm}
\hrule
\end{center} 
    \item \textbf{Final model selection.} On the basis of the comparative performance assessment in the Design Stage, pick the best-performing model and use that for the Analysis Stage. Start by re-training the model on the full pre-treatment sample using the hyperparameter(s) selected in the Design Stage.
    \item \textbf{Counterfactual forecasting.} For each unit $i$, forecast the post-treatment counterfactual outcome $\widehat{\Y}_{i,t_0+k|t_0}(0)$. In case of a large-scale shock affecting important covariates, either rely solely on their past lags or repeat steps 1--6 to forecast $\widehat{\bb{X}}_{i,t_0+k|t_0}(0)$ and use these values to improve the forecast of $\widehat{\Y}_{i,t_0+k|t_0}(0)$. Leverage Explainable Artificial Intelligence tools to enhance the model's explainability and transparency of the estimated counterfactual.
    \item \textbf{Estimation of treatment effects.} For each unit $i$, estimate the individual treatment effect as in Equation (\ref{eqn:hat_mcite_1}) by taking the difference between the observed post-treatment outcome $\Y_{i,t_0+1}$ and the ML-generated potential outcome $\widehat{\Y}_{i,t_0+1|t_0}(0)$. For $k\geq 2$, if the selected ML algorithm in Step 4 of the Design Stage is linear, estimate $\hat{\tau}_{t_0+k}$ as in Equation (\ref{eqn:hat_mcite_k_lin}), otherwise use Equation (\ref{eqn:hat_mcite_k_nonlin}). Estimate other causal estimands, such as the ATE from Equation (\ref{eqn:hat_ate_cate}) or the temporal ATE from Equation (\ref{eqn:hat_ate_avg}), by aggregating the individual estimates.
    \item \textbf{Treatment effect heterogeneity.} To uncover heterogeneity, data-driven CATEs can be estimated as in Equation (\ref{eqn:hat_ate_cate}) via a regression tree analysis with the individual treatment effects as the outcome variable and a set of variables potentially associated with treatment effect heterogeneity.
    \item \textbf{Inference.} Compute standard errors for the ATE and CATEs via block-bootstrap.
\end{enumerate}
\end{MyBox}
\vspace{5mm}
\end{spacing}
\end{footnotesize}

Our panel CV approach adapts time series CV based on expanding training windows to a panel setting.\footnote{Here we use an expanding window (as we are in a short-panel setting), but the procedure can also be implemented using a rolling window approach, which might be more appropriate in specific cases (e.g., with long panels or non-stationarity of the outcome).} The intuition is provided in Figure \ref{fig:pcv} and constitutes an adaptation from  \cite{Hyndman:Athanasopoulos:2021}. 

We start from a set of candidate algorithms $\mathcal{H} = (h_1, \dots, h_J)$, each having its own set of parameters $\theta^{(j)}$ and hyperparameters $\gamma^{(j)}$. For example, in the empirical application, we use four learners: LASSO, PLS, random forest and stochastic gradient boosting. For each learner, we carry out panel CV as detailed in Algorithm \ref{algorithm_PCV}. In short, the ML algorithms are trained repeatedly using an expanding window approach, with hyperparameters tuned at each step to minimize forecast errors. This sequential procedure ensures that, for each unit, there are no ``future'' observations in the training set and no ``past'' observations in the validation set. 
Note that the same procedure can be applied to forecast post-treatment values of important covariates affected by the intervention. We can assume that such covariates follow Equation (\ref{eqn:model_assumpt}) and use the ML algorithms and the panel CV routine to forecast counterfactual values of $\widehat{\bb{X}}_{i, t_0+k|t_0}(0)$, which we can then incorporate into the prediction of $\widehat{\Y}_{i,t_0+k|t_0}$ as in \citet{liu2020forecasting}.

\begin{algorithm}[h!]
\begin{algorithmic}[1]
\caption{Panel CV for one-step-ahead forecasting}
\label{algorithm_PCV}
\REQUIRE $\mathcal{H} = (h_1, \dots, h_J)$, $t_0$ 
\FOR{$j$ in $1:J$, where $J$ is the total number of candidate algorithms}
\FOR{$s$ in $1:t_0-1$}
\STATE Split the data into a training set $\mathcal{T} = \{\Y_{i,t},Z_{i,t} \}_{t = 1}^s$ and a validation set $\mathcal{V} = \{ \Y_{i,t}, Z_{i,t} \}_{t=s+1}$, where $Z_{i,t} = \{ \Y_{i,t-1}^{(p)}, \bb{X}^{(q)}_{i,t}\}$ is a set of features including past lags of the outcome and covariates that should be related to $\Y_{i,t}$.
\STATE Fit the model in the training set by optimizing a performance measure, e.g., minimizing MSE, 
$$\hat{\theta}^{(j)} = \argmin_{\theta}{ \sum_{i,t \in \mathcal{T}}(\Y_{i,t} - h_j(Z_{i,t}; \gamma^{(j)}, \theta^{(j)}))^2}$$ and compute the $1$-step-ahead forecast, i.e., $\E[\Y_{i,s+1}|Z_{i,s+1}] =  h_j (Z_{i,s+1}; \gamma^{(j)}, \hat{\theta}^{(j)})$. \STATE Tune hyperparameters in the validation set by optimizing the one-step-ahead forecast error,  $$\hat{\gamma}^{(j)} = \argmin_{\gamma} \sum_{i}(\Y_{i,s+1} - h_j(Z_{i,s+1}; \gamma^{(j)}, \hat{\theta}^{(j)}))^2$$
\STATE Repeat steps 3-5 for all $s$ up to $t_0-1$, the optimal values for the hyperparameters will be those that minimize the forecast errors across all horizons.
\ENDFOR
\STATE Repeat steps 1-6 for all $j$ and choose the algorithm in $\mathcal{H}$ which minimizes the one-step-ahead forecast error.
\ENDFOR
\end{algorithmic}
\end{algorithm}


To provide evidence about internal validity, we suggest running diagnostic checks (e.g., showing that the distribution of the pre-treatment forecasting errors is approximately Gaussian and centered around zero) as well as placebo tests, which have become a key device for assessing the credibility of research designs in observational settings \citep{eggers2024placebo}. Following \citet{Liu:Wang:Xu:2022}, panel placebo tests can be implemented by hiding one or more periods of observations right before the onset of the treatment and using a model trained on the rest of the pre-treatment periods to predict the untreated outcomes of the held-out period(s). If the identifying assumptions are valid, the differences between the observed and forecast outcomes in those periods should be close to zero.\footnote{If large discrepancies, i.e., forecasting errors, are found at this stage, this would indicate pre-intervention shocks or policy changes occurring between pre-treatment periods. These factors should be accounted for in the model to bridge the gap between observed and forecasted pre-treatment outcomes.} Given that we estimate unit-level treatment effects, we are also able to test whether most unit-level placebo differences are close to zero. The Design Stage thus ends with a battery of performance, diagnostic, and placebo tests.

The Analysis Stage starts with final model selection and training: on the basis of the comparative performance assessment in the Design Stage, pick the best-performing model, re-train it on the full pre-treatment sample (using the hyperparameter values obtained in the Design Stage), then use it to forecast counterfactual outcomes in the post-intervention period. Once that is done, treatment effects for each unit are given by the difference between the post-treatment observed data and the corresponding ML-generated counterfactual forecasts. The ATE is the average of the individual effects.

When dealing with a single post-intervention period, step 7 of the Analysis Stage is straightforward. However, when performing a multi-step-ahead forecast using a data-driven ML routine, one must take special care to avoid including post-treatment outcomes in the prediction. Multi-step-ahead forecast becomes more challenging when the selected ML algorithm is non-linear, as we cannot simply plug-in the forecasted value $\widehat{\Y}_{t_0+1|t_0}(0)$ into the already estimated model. As explained in Section \ref{subsec:estimators}, one solution is to re-estimate the potential outcome model for values of $k \geq 2$. Importantly, at the end of the panel CV procedure, we know the selected ML algorithm and can adjust our forecasting strategy accordingly. Specifically, we propose using Equation (\ref{eqn:direct_models}) to model post-intervention potential outcomes, as this approach has proven effective in identifying non-linear, multi-step-ahead causal effects, and estimate them with Equation (\ref{eqn:hat_mcite_k_nonlin}). From a practical standpoint, this requires repeating the panel CV for each step of the forecasting horizon. The detailed procedure is described in Algorithm \ref{algorithm_multi_forecast}.  

\begin{algorithm}[h!]
\begin{algorithmic}[1]
\caption{Panel CV for non-linear multi-step-ahead forecasting}
\label{algorithm_multi_forecast}
\REQUIRE $h^{win}$ (selected model after Algorithm \ref{algorithm_PCV}).
\STATE Compute $\widehat{\Y}_{i,t_0+1|t_0} = \widehat{h}^{win}(y_{i,t_0}^{(p)},\bb{x}_{i,t_0+1}^{(q)}) $
\IF{$h^{win}$ is a non-linear model and the forecasting horizon $k \geq 2$}
\FOR{$k = 2$}
\STATE Start from a set of candidate algorithms  $\mathcal{G} = (g_2^{(1)}, \dots, g_2^{(j)})$
\STATE Split the data into a training set $\mathcal{T} = \{\Y_{i,t_0}^{(p)},\bb{X}_{i,t_0}^{(q)} \}$ and a validation set $\mathcal{V} = \{ \widehat{\Y}_{i,t_0+1|t_0}, \bb{X}_{i,t_0+1}^{(q)} \}$.
\STATE Fit the model $g_2^{(j)}$ in the training set by optimizing a performance measure, e.g., minimizing MSE, 
$$\hat{\theta}^{(j)} = \argmin_{\theta}{ \sum_{i,t_0 \in \mathcal{T}}(\Y_{i,t_0} - g_2^{(j)}(\Y_{t_0-1}^{(p)}, \bb{X}_{i,t_0}^{(q)}; \gamma^{(j)}, \theta^{(j)}))^2}$$ and compute the $1$-step-ahead forecast based on $g_2$, i.e., $g_2^{(j)} (\Y_{t_0}^{(p)}, \bb{X}_{i,t_0+1}^{(q)}; \gamma^{(j)}, \hat{\theta}^{(j)})$.
\STATE Tune hyperparameters in the validation set by optimizing the one-step-ahead forecasting error,  $$\hat{\gamma}^{(j)} = \argmin_{\gamma} \sum_{i}(\widehat{\Y}_{i,t_0+1|t_0} - g_2^{(j)}(\Y_{t_0}^{(p)}, \bb{X}_{i,t_0+1}^{(q)}; \gamma^{(j)}, \hat{\theta}^{(j)}))^2.$$
\STATE Choose the candidate algorithm in $\mathcal{G}$ which minimizes the one-step-ahead forecast error, retrain the model including the validation set and compute the one-step-ahead forecast $\widehat{\Y}_{t_0+2|t_0} = \widehat{g}_2(\widehat{\Y}_{t_0+1|t_0}, \Y_{t_0}^{(p)}, \bb{X}_{t_0+2}^{(q)})$.
\STATE Repeat steps 4-8 for all $k > 2$
\ENDFOR
\ENDIF
\end{algorithmic}
\end{algorithm}

During the counterfactual forecasting step, especially when opting for more complex and less transparent techniques, users should consider leveraging innovations in Interpretable Machine Learning and Explainable Artificial Intelligence \citep{molnar2020interpretable}, such as model-agnostic Shapley Additive Explanations (SHAP) \citep{lundberg2017unified}, to enhance explainability and transparency of the counterfactual building process \citep{Abadie:2021}.
\frenchspacing

After estimating individual effects, data-driven CATEs can be computed via a regression tree analysis. Specifically, this approach uses the estimated treatment effects as the outcome variable, regresses them on many potentially associated variables, and lets the algorithm pick the main predictors and their critical thresholds. The resulting tree reports the group-average treatment effects for all units in each terminal node. As for causal trees and forests \citep{Athey:Imbens:2016, Wager:Athey:2018}, this data-driven search for heterogeneity of causal effects removes a major degree of discretion because the researcher can only select the set of covariates that can be used by the tree to build the subgroups. However, the two approaches differ regarding both purpose and implementation: our data-driven technique is a post-estimation approach aimed at automatically recovering and visualizing the relevant heterogeneity dimensions, while causal trees and forests are counterfactual methods for the direct estimation of heterogeneous treatment effects, which they achieve by leveraging control units. Moreover, we are not interested in the out-of-sample performance of the regression tree, but in retrieving CATEs for the entire population of interest. To this end, there is no need to split the sample into training and testing sets or to prune the tree by adjusting the complexity parameter. However, it may be necessary to set a pre-specified minimum node size to preserve the interpretability of the resulting tree.

Finally, standard errors and confidence intervals for the ATE and CATEs are estimated through the block-bootstrap approach described in Supplemental Appendix \ref{app:bootstrap}.\footnote{In staggered adoption settings, uncertainty quantification with block-bootstrapping can be extended to other estimands, such as group-time average treatment effects. See the replication exercise in Supplemental Appendix \ref{app:replication}.}

\subsection{Simulation study}
\label{subsec:simulation}
To investigate the performance of the MLCM in detecting average treatment effects in panel datasets, we performed an extensive simulation study using different data-generating processes (linear and non-linear) and various combinations of pre- and post-intervention periods.
We generated $1,000$ panel datasets, each consisting of $400$ units and $T = 7$ or $12$ time periods, according to the following two models,
\begin{align*}
    \Y_{i,t}(0) & = \phi \Y_{i,t-1}(0) + \bb{X}_{i,t-1}\boldsymbol{\beta} + \epsilon_{i,t} & \text{(Linear)}\\
    \Y_{i,t}(0) & = \sin \left\{ \phi \Y_{i,t-1}(0) + \bb{X}_{i,t-1}\boldsymbol{\beta} \right\} + \epsilon_{i,t} & \text{(Non-Linear)}
\end{align*}
where: the error term $\epsilon_{i,t}$ is generated from a Normal distribution with standard deviation $\sigma_{\epsilon} = 2$, i.e.,  $\epsilon_{i,t} \sim N(0,2)$, $\phi = 0.8$ is the autoregressive coefficient, $\boldsymbol{\beta} = (\beta^{(1)}, \dots, \beta^{(11)})'$ is a $m \times 1$ vector of coefficients and $\bb{X}_{i,t-1} = \left(\X_{i,t-1}^{(1)}, \dots, \X_{i,t-1}^{(11)} \right)$ is a $1 \times m$ vector of $11$ predictors, both continuous and categorical, also containing interaction terms and correlated regressors. We allowed the covariates to vary in time and across units in the dataset by adding, respectively, a random term $\nu_t$ (which, as will become clear later, varies for different covariates) and a random term $u_i \sim N(1,1)$. The latter term adds variability (and thus, heterogeneity) between the units. In particular, the covariates are generated as follows: $\X_{i,t}^{(1)}  = 0.1 t+u_i+\nu_t^{(1)} ,  \nu_t^{(1)}  \sim N(0,1)$, $\X_{i,t}^{(2)}  = 0.1 t+u_i+\nu_t^{(2)} ,  \nu_t^{(2)}  \sim N(0,0.2)$, $\X_{i,t}^{(3,4,5)}  = u_i+\nu_t^{(3,4,5)} ,  \nu_t^{(3,4,5)}  \sim MVN(0, \Sigma)$, $\X_{i,t}^{(6)}  = u_i-\nu_t^{(6)},  \nu_t^{(6)}  \sim N(0,1)$, $\X_{i,t}^{(7)} = \left(0.1 t + \nu_t^{(1)} \right)^2 + u_i + \nu_t^{(7)},  \nu_t^{(7)}  \sim N(0,0.2)$, $\X_{i,t}^{(8)} \in \{0,1\}$, $\X_{i,t}^{(9)} \in \{1,2,3\}$, $\X_{i,t}^{(10)} = \X_{i,t}^{(3)} \cdot \X_{i,t}^{(9)}$, $\X_{i,t}^{(11)} = \X_{i,t}^{(2)} \cdot \X_{i,t}^{(8)}$.\footnote{The variance-covariance matrix of the multivariate normal distribution used to generate covariates 3--5 is set to $\Sigma = \begin{bmatrix}
    1 & .5 & .7 \\
    .5 & 1 & .3 \\
    .7 & .3 & 1 \\
\end{bmatrix}$. Notice that, to put the ML algorithms under further stress, the covariance between $\X_{i,t}^{(3)}$ and $\X_{i,t}^{(5)}$ is $0.7$ so the two variables are highly correlated but $\X_{i,t}^{(3)}$ is ten times more important (in terms of coefficient) than $\X_{i,t}^{(5)}$.}
These choices regarding the data-generating process for the covariates included in our simulation study are driven by the goal of mirroring real-world empirical settings where the methodology might be applied. In addition, to better reflect a typical real-world scenario where only a subset of covariates is relevant, we set certain coefficients to zero: specifically, $\beta^{(1)}$, $\beta^{(8)}$, and $\beta^{(9)}$. The remaining coefficients are generated as follows: $\beta^{(2)} = \beta^{(6)} = \beta^{(10)} = 2$, $\beta^{(3)} = \beta^{(7)} = 1$, $\beta^{(4)} = 2.5$, $\beta^{(5)} = 0.1$ and $\beta^{(11)} = 1.5$. We also remark that, for computational reasons, in this simulation study we focus on a low-dimensional set of covariates. However, the MLCM is particularly well-suited for data-rich environments and sparse settings with many more covariates. Therefore, the simulation results provided here should be interpreted as conservative evidence regarding the performance of the MLCM. Finally, we assume exogeneity of the predictors in both the pre- and post-intervention periods (third part of Assumption \ref{assumpt:no_anticipation}). Table \ref{tab:sim_ex_data} below provides an overview of the generated datasets. 
\vspace{-8pt}
\begin{table}[h!]
\centering
\caption{First 9 observations from one of the datasets generated during the simulation study}
\label{tab:sim_ex_data}
\begin{tabular}{cccccccccccccc} \hline 
Time & ID & $\Y$ & $\X^{(1)}$ & $\X^{(2)}$ & $\X^{(3)}$ & $\X^{(4)}$ & $\X^{(5)}$ & $\X^{(6)}$ & $\X^{(7)}$ & $\X^{(8)}$ & $\X^{(9)}$ & $\X^{(10)}$ & $\X^{(11)}$ \\ \hline
1             & 1           & 16.57      & 1.71        & 0.9         & 1.38        & 3.61        & 3.03        & 0.42        & 1.09        & 0           & 1           & 1.38         & 0            \\
2             & 1           & 43.62      & 1.62        & 1.53        & 2.4         & 3.33        & 3.73        & 1.14        & 1.24        & 0           & 3           & 7.19         & 0            \\
3             & 1           & 73.62      & 2.83        & 1.71        & 3.13        & 3.44        & 3.99        & 3.89        & 3.37        & 1           & 1           & 3.13         & 1.71         \\
4             & 1           & 89.04      & 1.16        & 1.91        & 1.53        & 3.35        & 2.25        & 1.14        & 1.25        & 1           & 3           & 4.58         & 1.91         \\
5             & 1           & 170.17     & 2.61        & 1.62        & 1.99        & 1.47        & 3.46        & 1.23        & 3.79        & 0           & 2           & 3.98         & 0            \\
6             & 1           & 153.05     & 1.43        & 1.63        & 1.3         & 3.3         & 3.2         & 1.22        & 1.65        & 0           & 1           & 1.3          & 0            \\
7             & 1           & 133.43     & 0.3         & 1.8         & 0.46        & 1.43        & 3.38        & 1.32        & 1.14        & 1           & 3           & 1.38         & 1.8          \\
1             & 2           & 16.88      & 1.42        & 1.14        & 0.23        & 2.93        & 3           & 0.6         & 2.11        & 1           & 3           & 0.7          & 1.14         \\
2             & 2           & 45.33      & 1.66        & 1.75        & 2.94        & 3.2         & 3.57        & 0.56        & 1.13        & 0           & 2           & 5.87         & 0    \\ \hline         
\end{tabular}
\end{table}

At $t_0+1$ (corresponding to T = $5$ in Table \ref{tab:sim_ex_data}), we included a fictional intervention that increases the outcome for each unit by $2$ standard deviations at time $t_0+1$, $1.5$ standard deviations at time $t_0+2$, and $1$ standard deviation at time $t_0+3$. In other words, the intervention has a decreasing effect over time. We made this choice because adding a unit-specific component in the covariates generates heterogeneity; therefore, the scale of each $\Y_i$ varies across the $i$’s. We measured the performance of the MLCM in terms of both the bias of the estimated effect from the true impact and the interval coverage. Also note that, under this setup, the number of pre-intervention periods is $t_0 = 4$ when $T = 7$ and $t_0 = 9$ when $T = 12$. As estimators, we employ four different ML algorithms - LASSO, Partial Least Squares, random forest, and stochastic gradient boosting.

The results are summarized in Table \ref{tab:sim_res} below and show that, across all post-intervention horizons, the MLCM achieves a very low bias. This holds true for both linear and non-linear model specifications. The bias tends to decrease when the number of pre-intervention time periods increases, as more information is present in the data. Nevertheless, the bias at the shortest time period is still very low, which reinforces our belief that the MLCM can be effectively used for short panels. For instance, under the linear specification and $t_0 = 4$, the relative bias increases from $0.2\%$ measured at $t_0+1$ to $0.9\%$ measured at $t_0+3$. However, by increasing the number of pre-intervention time points, the bias drops at $0.4\%$ even at $t_0+3$. We also observe that the relative bias under a linear model specification is much lower than that under the non-linear specification, which was expected since non-linearities are typically more challenging to detect. In any case, the bias under a non-linear model specification is still low (the maximum bias is $4.9\%$ and is measured in the most challenging scenario, i.e., at the third time horizon when the pre-intervention series is very short, $t_0 = 4$).

\begin{table}[h]
\centering
\caption{Simulation results for the linear and non-linear model specifications}
\label{tab:sim_res}
\scalebox{0.83}{
\begin{tabular}{cccccccccccc} \hline 
\multicolumn{12}{c}{1st Post-intervention period ($t_0+1$)}  \\ \hline
& & & \multicolumn{4}{c}{Linear} & & \multicolumn{4}{c}{Non-Linear} \\ \cline{4-7} \cline{9-12}
Pre-int.           & Bootstrap   &  & True ATE  & Bias  & Rel. Bias  & Coverage     &      & True ATE  & Bias  & Rel. Bias  & Coverage \\
4 & block   & & 74.24 & 0.14 & 0.002 & 0.94 & & 4.06 & 0.10 & 0.024 & 0.95 \\ 
9 & block & & 93.56 & 0.10 & 0.001 & 0.95 & & 4.14 & 0.10 & 0.023 & 0.94 \\ \hline
\multicolumn{12}{c}{2nd Post-intervention period ($t_0+2$)}    \\ \hline                                   & & & \multicolumn{4}{c}{Linear} & & \multicolumn{4}{c}{Non-Linear} \\  \cline{4-7} \cline{9-12}                                     
Pre-int.           & Bootstrap   &  & True ATE  & Bias  & Rel. Bias  & Coverage     &      & True ATE  & Bias  & Rel. Bias  & Coverage \\
4 & block   & & 55.68 & 0.22 & 0.004 & 0.92 & & 3.05 & 0.10 & 0.031 & 0.90 \\ 
9 & block & & 70.17 & 0.15 & 0.002 & 0.91 & & 3.10 & 0.09 & 0.031 & 0.92 \\ \hline
\multicolumn{12}{c}{3rd Post-intervention period ($t_0+3$)}    \\ \hline                                  & & & \multicolumn{4}{c}{Linear} & & \multicolumn{4}{c}{Non-Linear} \\  \cline{4-7} \cline{9-12}                                      
Pre-int.           & Bootstrap   &  & True ATE  & Bias  & Rel. Bias  & Coverage     &      & True ATE  & Bias  & Rel. Bias  & Coverage \\
4 & block   & & 37.12 & 0.33 & 0.009 & 0.93 & & 2.03 & 0.10 & 0.049 & 0.93 \\ 
9 & block & & 46.78 & 0.19 & 0.004 & 0.92 & & 2.07 & 0.09 & 0.046 & 0.93 \\ 
\hline
\multicolumn{12}{p{1.2\textwidth}}{\justify \begin{footnotesize} \justify \textit{Notes.} These results are obtained under the following scenario: $N = 400$ units, $\phi = 0.8$, and $\sigma_u = 1$, where $\sigma_u$ is the standard deviation of the random heterogeneity term $u_i$. In this table, ``Pre-int.'' denotes the number of pre-intervention periods. \end{footnotesize}}   
\end{tabular}}
\end{table}

Similar observations can be made for the interval coverage, which was estimated based on $1,000$ block-bootstrap iterations. At the first time horizon after the intervention, the interval coverage is close or equal to the nominal $95\%$ level for both linear and non-linear model specifications. Then, it tends to slightly decrease as we move further away from the intervention, but remains close to the nominal level for both linear and non-linear processes. 


Overall, these numerical studies demonstrate that the MLCM can achieve forecast unbiasedness and that the coverage of the estimator is high even when considering short pre-intervention windows. The simulation results also suggest that when we are interested in the estimation of causal effects $k$-step ahead from the treatment, having a greater number of pre-intervention periods further mitigates the bias, both under linear and non-linear data-generating processes. 

In Supplemental Appendix \ref{app:simulation}, we provide many additional results that show that these key findings remain largely unchanged if we consider alternative data-generating processes for both the outcome variable and the covariates, and if we reduce the number of available units. Notably, when  $\phi$ is set to $1.2$ (cf. Table \ref{tab:n400phi12u1}), rather than to $0.8$ as in Table \ref{tab:sim_res}, i.e., when moving from a stationary to a non-stationary data-generating process with an explosive autoregressive coefficient for the lag of the outcome variable, the MLCM exhibits very similar performance. 
\section{Empirical application}
\label{sec:application}
The main empirical illustration provided below is an original study on the educational effects of the COVID-19 crisis, thus focusing on a simultaneous treatment affecting all available units. In Supplemental Appendix \ref{app:replication}, to demonstrate how to leverage the MLCM in settings where there are untreated units, but there may be violations of the no-interference assumption, we revisit the application on minimum wage with staggered treatment adoption in \citet{callaway2021difference} without using their control group. Both empirical illustrations rely on short panel datasets.
\subsection{Background}
\label{subsec:background}
In the aftermath of the COVID-19 pandemic, the inequality legacy of this unprecedented crisis has emerged as an issue of great policy relevance. The available evidence on income inequality documents a decrease driven by short-run and temporary government compensation policies, which were mostly targeted at the poorest segments of the population \citep{Stantcheva:2022}. Concerning education, instead, recent micro-level evidence \citep{Agostinelli:Doepke:Sorrenti:Zilibotti:2022, Carlana:Ferrara:Lopez:2023}  shows that school closures had a large, persistent, and unequal effect on learning. The educational gaps caused by the school closures may, in turn, permanently affect the lifetime income possibilities of the current generation of students, with vast repercussions on future inequalities \citep{Werner:Woessmann:2023}. Therefore, it is through the human capital channel that the inequality effects of the pandemic may eventually appear in the medium and long run. It follows that to anticipate longer-term consequences on income inequality and territorial disparities, it is necessary to gauge the magnitude of educational losses and their distribution across areas.

To our knowledge, there is no granular evidence on the geography of the education effects of the COVID-19 crisis. This is mainly due to identification challenges caused by the sudden spread of the pandemic across the world, which resulted in the absence of a suitable untreated group. Italy is an important case study, as it ranks among the hardest-hit countries and was the first Western country to impose a strict nationwide lockdown. During the lockdown that began on March 9, 2020, schools across the entire national territory were closed and remained so until the end of the school year. This resulted in Italy ranking among the OECD countries with the highest number of weeks of school closures and distance learning \citep{Battisti:Maggio:2023}.  In this setting, we leverage the MLCM with non-staggered adoption, given that the treatment---the COVID-19 shock---simultaneously affected all units.\footnote{We refer to the treatment variable as COVID-19 ``shock'', which encompasses both the pandemic and the containment policies, as we aim to capture the total effect of the pandemic, i.e., its direct and indirect effects on educational outcomes. In line with previous literature, we consider the restrictive measures, including the lockdown and the school closures, as a manifestation of the pandemic, not as a separate treatment. COVID-related learning losses can originate from many pandemic-induced channels, such as school closures and distance learning, prolonged absence from school due to one or more infections, absence due to parents’ concerns about their children’s health, and many other transmission mechanisms. At the aggregate level, only the overall impact of the pandemic remains discernible.}

\subsection{Data and implementation}
\label{subsec:data}
We employ LLM yearly data covering the period from 2013 to 2020. We cover all Italian LLMs except some of the smallest ones for which the education data are unavailable due to privacy protection, for a total of $579$ LLMs ($95\%$ of Italian LLMs and over $99\%$ of the Italian population). The dependent variable is the standardized math test score of fifth-grade students referred to the school year started in 2020. Following the approach by \citet{Carlana:Ferrara:Lopez:2023}, the math score is standardized with respect to its pre-pandemic mean. This way, the detected treatment effects can be interpreted as deviations from the pre-COVID baseline. We focus on younger students for two reasons: i) learning disruptions earlier in life typically have longer-lasting and more severe effects, so younger children may have been more heavily impacted \citep{Stantcheva:2022}; ii) primary schools were excluded from the heterogeneous school closures involved in the tier system of regional restrictions implemented by the Italian government since the onset of the second wave (November 2020), so the treatment is the same for all units (Assumption \ref{assumpt:sutva}).  The policy relevance of the math outcome is illustrated by the fact that the most recent report by the OECD Program for International Student Assessment (PISA), published in December 2023, reported that math scores dropped globally in the last few years, making headlines around the world.\footnote{See \href{https://www.nytimes.com/2023/12/05/us/math-scores-pandemic-pisa.html}{\textcolor{blue}{here}} for coverage of the news from the New York Times.}

The initial pre-treatment information set includes over $150$ variables (see Table \ref{tabapp:vardef} in Supplemental Appendix \ref{app:application} for a detailed description of the variables). In this set of covariates, we included the first three lags of all the predictors as covariates and three lags of the outcome variable. This implies that we collapse the original 2013–2020 dataset into a dataset covering the period 2017–2020.

The implementation process for this application is reported in Box \ref{box:application}, while the outcomes of the empirical analysis are reported in Section \ref{subsec:results}.

\vspace{0.2cm}

\begin{spacing}{1}
\begin{footnotesize}
\begin{MyBox}{\hspace{2cm}{Box \ref{box:application}: Estimating the local impact of the COVID-19 shock on education}}{2}
\label{box:application}
\begin{center}
   \hspace{0.75cm} \textbf{Preliminary} 
    \vspace{2mm} 
    \hrule
\end{center} 
\vspace{1mm}
\begin{enumerate}[start=0]
    \item \textbf{Data splitting.} We split the full 2017-2020 dataset into two subsets according to the treatment date: 2017-2019 to be used in the Design Stage; 2020 to be used in the Analysis Stage.
\vspace{2mm} 
\hrule
\begin{center}
\textbf{Design Stage} 
\vspace{2mm}
\hrule
\end{center} 
    \item \textbf{Algorithm selection.} We select four supervised ML algorithms (the same set used for the simulation of Section \ref{subsec:simulation}): 1) LASSO; 2) Partial Least Squares;  3) stochastic gradient boosting; 4) random forest. As we are agnostic about the functional form of the underlying data-generating process, we opt for a mix of non-linear and linear models.
    \item \textbf{Principled input selection.} We build an initial LLM dataset with over $150$ predictors on the basis of literature insights. From this dataset, we then keep only the most important predictors selected by a preliminary random forest run on the pre-treatment data (see step below).
    \item \textbf{Panel cross-validation.} For each algorithm, we tune hyperparameters via panel CV, involving iterative estimation on two different training-testing pairs of pre-COVID datasets: i) 2017 training; 2018 testing; ii) 2017-2018 training, 2019 testing.
    \item \textbf{Performance assessment.} We assess average performance metrics for the four algorithms by comparing average forecasted vs. actual outcomes on the 2018–2019 held-out test data. We then compare the performance of the different MLCM versions.
    \item \textbf{Diagnostic and placebo tests.} We first check the average distribution of errors with the best-performing model for the 2018-2019 testing sets and then show the map of the unit-level placebo temporal average treatment effects in the pre-COVID period.    
\vspace{3mm} 
\hrule
\begin{center}
    \textbf{Analysis Stage} 
    \vspace{2mm}
    \hrule
\end{center} 
    \item \textbf{Final model selection.} On the basis of the comparative performance assessment, we pick the best-performing algorithm (random forest), with its best configuration, and re-train it on the full 2017–2019 sample using the hyperparameters cross-validated in the Design stage.
    \item \textbf{Counterfactual forecasting.} We apply the final model estimated in Step 6 and forecast, for each LLM $i$, the counterfactual outcome $\widehat{\Y}_{i,t_0+1|t_0}(0)$.
    \item \textbf{Estimation of treatment effects.} For each LLM $i$, we estimate the individual treatment effect by taking the difference between the observed post-COVID outcome $\Y_{i,t_0+1}$ and the ML-generated potential outcome $\widehat{\Y}_{i,t_0+1|t_0}(0)$.
    \item \textbf{Treatment effect heterogeneity.} We estimate data-driven CATEs via a regression tree analysis with the individual treatment effects as the outcome and a host of potentially relevant predictors associated with the heterogeneity of the education effects.
    \item \textbf{Inference.} We estimate standard errors for the ATE and CATEs via block-bootstrapping by performing $1,000$ bootstrap replications of Steps 6 to 9.
\end{enumerate}

\end{MyBox}
\end{footnotesize}
\end{spacing}
\vspace{10pt}

We use a data-driven approach to restrict the information set. More specifically, we follow \citet{Athey:Wager:2019} and \citet{Basu:Kumbier:Brown:Yu:2018} and apply a pilot random forest on the pre-treatment data to pick up a subset of the most important predictors according to the importance ranking produced by the forest.\footnote{For this preliminary operation, we use default hyperparameter settings that typically perform well with random forests \citep{Athey:Wager:2019}. See Figure \ref{appfig:rf_varimp} for the variable importance ranking of the preliminary forest.} In order to select the precise number of relevant predictors in a data-driven manner, we include this number as an additional parameter in the subsequent panel CV routine which we employ to tune the hyperparameters of all the selected ML algorithms (LASSO, Partial Least Squares, stochastic gradient boosting, and random forest
).\footnote{For all algorithms, we tune the subset of most predictive variables (10, 20 or 30) to be included in the analysis according to the preliminary forest. For LASSO, we tune the penalty parameter (the values from 0.1 to 0.9 with an increase of 0.1 in each run); for PLS, we tune the number of components (all possible values from 1 to the number of variables). For boosting, we tune the number of trees (1,000 or 2,000), the maximum depth of each tree (1 or 2), the minimum number of observations in terminal nodes (5 or 10) and the learning rate (0.001, 0.002 or 0.005); for random forest, we tune the number of variables randomly sampled as candidates at each split (1/2, 1/3 or 1/4 of the number of variables), and use a fixed number of 1,000 trees. All candidate hyperparameter values are tested on each testing set.}

Note that in this application, due to the ubiquitous nature of this shock, we do not invoke the third part of Assumption \ref{assumpt:no_anticipation} (exogeneity of post-treatment covariates) and only include pre-treatment values of time-varying variables and time-invariant predictors in the information set. Assumption \ref{assumpt:no_anticipation} is satisfied because the pandemic was completely unanticipated. Regarding Assumption \ref{assumpt:model}, given the unprecedented disruptions brought about by the pandemic and based on institutional and subject matter knowledge, we deem it plausible that any divergence from the forecasted counterfactual in educational outcomes can be attributed exclusively to the COVID-19 shock, and not to other concomitant shocks or policies.\footnote{There were no educational reforms at the primary school level that could have influenced the standardized math test scores of fifth-grade students in the period under scrutiny.} Finally, Assumption \ref{assumpt:additional_models} is not needed for identification, as we only estimate one-step-ahead causal effects here.

\subsection{Results}
\label{subsec:results}

\subsubsection{Design Stage}
\label{subsubsec:design}

Table \ref{tab:placebo} below reports the average performance—in terms of panel CV MSE—of the four selected ML algorithms in forecasting the standardized math test score of fifth-grade students across the pre-treatment test sets (2018 and 2019).\footnote{A replication notebook of the main results reported in this section is publicly available \href{https://marclet.github.io/MLCM-Replication-Notebook/}{\textcolor{blue}{here}}.} 

The best-performing MLCM version is the one using the random forest with twenty variables and 1/3 of them, i.e., six, as candidates at each split. Overall, the fully non-linear random forest and boosting fare better than the linear ML methods, suggesting the presence of non-linearities in the data-generating process. Moreover, the non-linear methods outperform LASSO and Partial Least Squares in all cases, and their performance is stronger in the presence of very few variables and less deteriorated by the inclusion of variables with poor predictive power (cf. Table \ref{tabapp:performance} in Supplemental Appendix \ref{app:application}). See Table \ref{tabapp:rf_included_vars} in Supplemental Appendix \ref{app:application} for a detailed description of the 20 variables with the highest importance score attributed by the preliminary forest).

\begin{table}[H]
\centering
\caption{Performance and placebo estimates}
\label{tab:placebo}
\begin{tabular}{lcl} \hline
\multicolumn{2}{l}{\textbf{Panel A – Performance}}         &  \\ \hline
\textbf{MLCM   using:}        & \textbf{Panel CV-MSE}               &  \\
LASSO                         & 0.4169                     &  \\
Partial Least Squares         & 0.4044                     &  \\
Boosting                      & 0.3909                     &  \\
Random Forest                 & 0.3902                     &  \\ \hline
\multicolumn{2}{l}{\textbf{Panel   B – Placebo estimates with Random Forest}} &  \\ \hline 
\textbf{Time   period}        & \textbf{Placebo ATE}       &  \\
2018                          & 0.0554                     &  \\
                              & (-0.1182 ; 0.2765)         &  \\
2019                          & 0.0254                     &  \\
                              & (-0.0273 ; 0.0728)         &  \\
Temporal average              & 0.0404                     &  \\ \hline
\end{tabular}
\vspace{0.1cm}

{ \justify \begin{footnotesize} 
\begin{spacing}{1.0}
\textit{Notes}: The panel CV procedure has selected the 20 most predictive variables from the dataset for all methods. In addition, for boosting, it has selected 2,000 trees, 2 as the maximum depth of each tree, 5 as the minimum number of observations in terminal nodes and 0.005 for the learning rate. The other selected hyperparameters are: 0.3 as the penalty parameter (LASSO), 9 as the number of components (PLS), and 6 for the number of variables randomly sampled as candidates at each split (random forest). Lower and upper bounds of placebo ATE estimates refer to the 95\% confidence intervals and are reported within brackets. 
\end{spacing}
\end{footnotesize} \par}
\end{table}

Panel CV on pre-treatment periods implicitly enables in-time placebo analysis along the lines of \citet{Bertrand:Duflo:Mullainathan:2004} and \citet{Liu:Wang:Xu:2022}. In-time placebos are performed on the same pool of treated units where we ``fake'' that the treatment occurred at time $t_0-1$ and use only information up to $t_0-1$ to forecast the counterfactuals at time $t_0$. Since we know what the real values at time $t_0$ are, we can shift the treatment date artificially backward in time to assess the difference with the main estimates and evaluate the forecasting accuracy and unbiasedness of our estimator. As reported in the table, the placebo ATEs are indistinguishable from zero and statistically insignificant both in 2018 (ATE: 0.0554, with 95\% confidence intervals (–0.1182; 0.2765) and 2019 (ATE: 0.0254, with 95\% confidence intervals (–0.0273; 0.0728).\footnote{The counterfactual estimates exhibit substantial similarity regardless of whether a preliminary random forest or an alternative preliminary procedure is employed, such as boosting or selecting covariates based solely on the highest absolute correlation coefficients (e.g., the correlation between the counterfactual estimates and those obtained using default hyperparameter values is 0.9985). Detailed results are available upon request.}

\begin{figure}[H]
    \centering    
    \includegraphics[scale = 0.55]{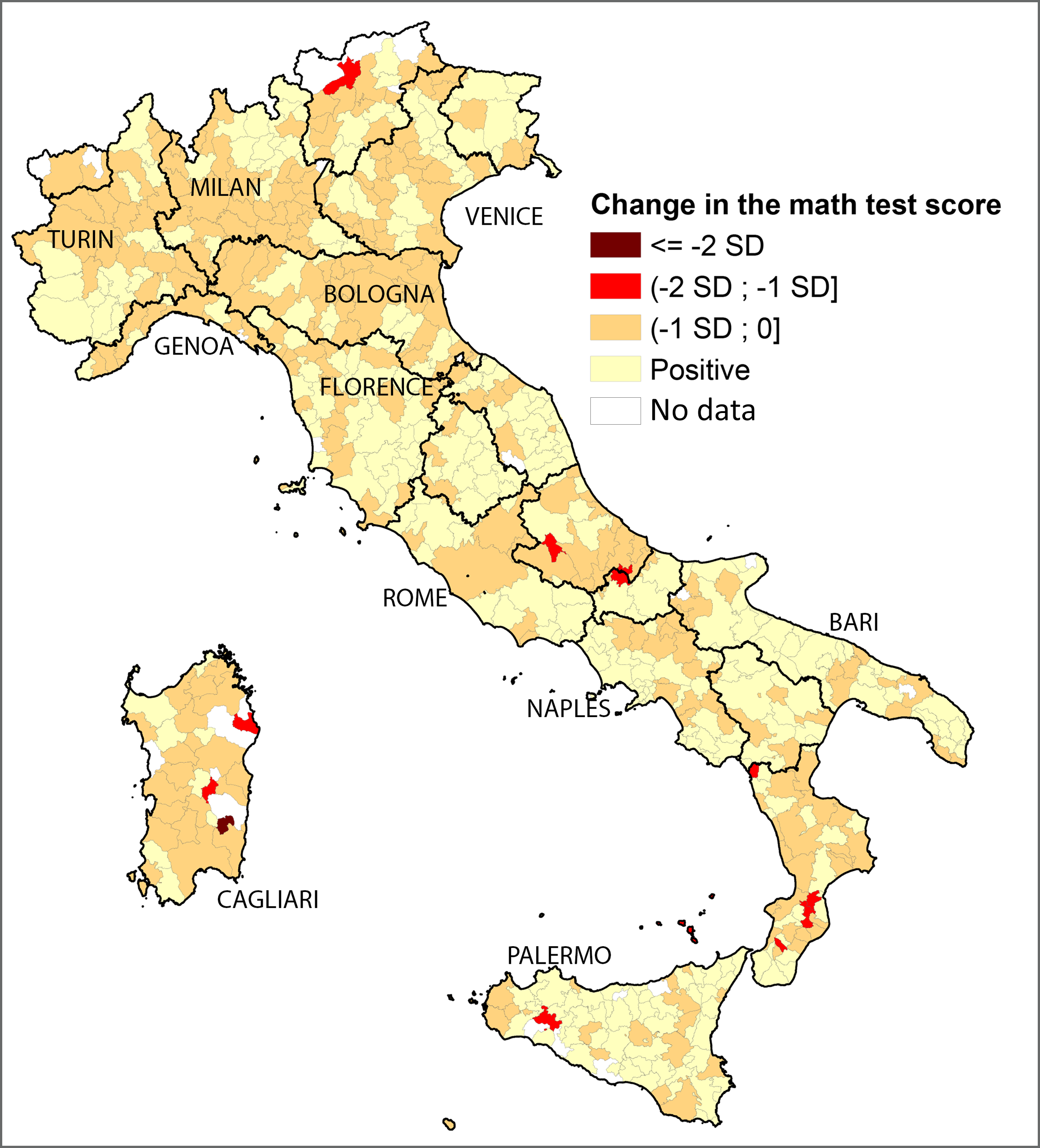}
    \caption{Unit-level panel placebo test—Temporal average `treatment' effects. \textit{Notes:} the standard deviation (SD) refers to the standardized pre-COVID mean. Estimation via the MLCM with random forest.}
    \label{fig:placebo}
\end{figure}

In addition, the unit-level placebo map reported in Figure \ref{fig:placebo} depicts the temporal average (for the years 2018 and 2019) individual `treatment' effects estimated with the best-performing ML algorithm, random forest, and shows that almost all LLMs exhibit no trace of significant differences between the forecasted and observed math scores in the years before the COVID-19 pandemic. In particular, only 2\% of the LLMs report a difference lower than –1 standard deviation.\footnote{Figure \ref{appfig:yearly_placebos} in Supplemental Appendix \ref{app:application} presents the placebo maps for the single pre-treatment years (2018 and 2019). In these instances as well, few LLMs exhibit a difference lower than –1 standard deviation between the forecasted and observed math scores prior to the pandemic (4.3\% of LLMs in 2018 and 5.7\% of LLMs in 2019). Furthermore, the LLMs with extreme values tend to be the smallest ones, which are also the most volatile.
}

\begin{figure}[H]
    \centering
    \includegraphics[scale = 0.225]{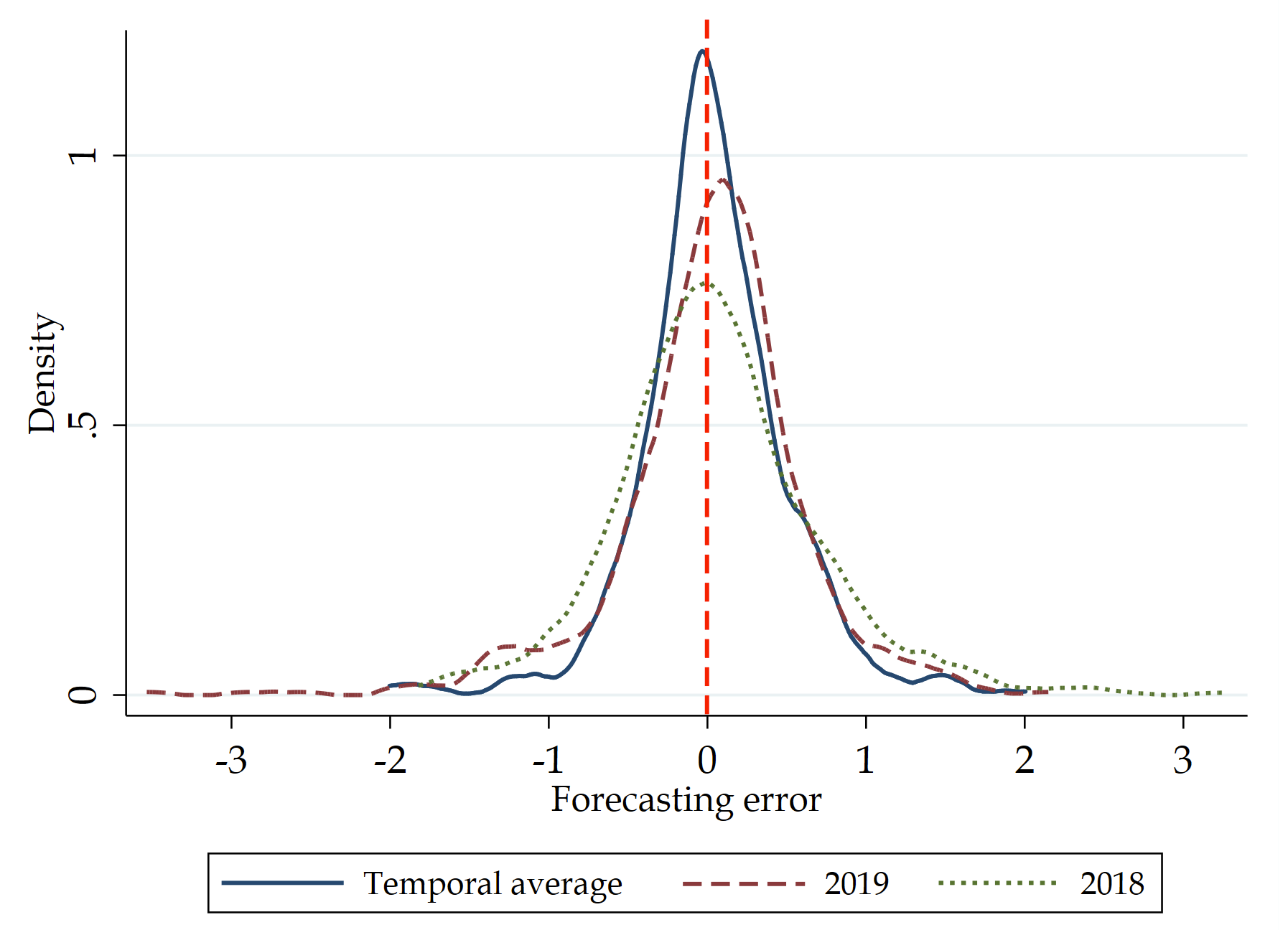}
    \caption{Distribution of the forecasting error. \textit{Note}: Forecasting errors of the MLCM estimated with random forest.}
    \label{fig:placebo_forecast_error}
\end{figure}

Finally, Figure \ref{fig:placebo_forecast_error} reports a diagnostic test for the performance of the MLCM with random forest. The figure illustrates that the distribution of the placebo forecasting errors is approximately normal and centered around zero, not only for the temporal average but also for the single pre-treatment years. This demonstrates that the ATE estimator is nearly unbiased even when applied on very short panels and without relying on the use of post-treatment information.

\subsubsection{Analysis Stage}
\label{subsubsec:analysis}
\begin{figure}[H]
    \centering
    \includegraphics[scale = 0.55]{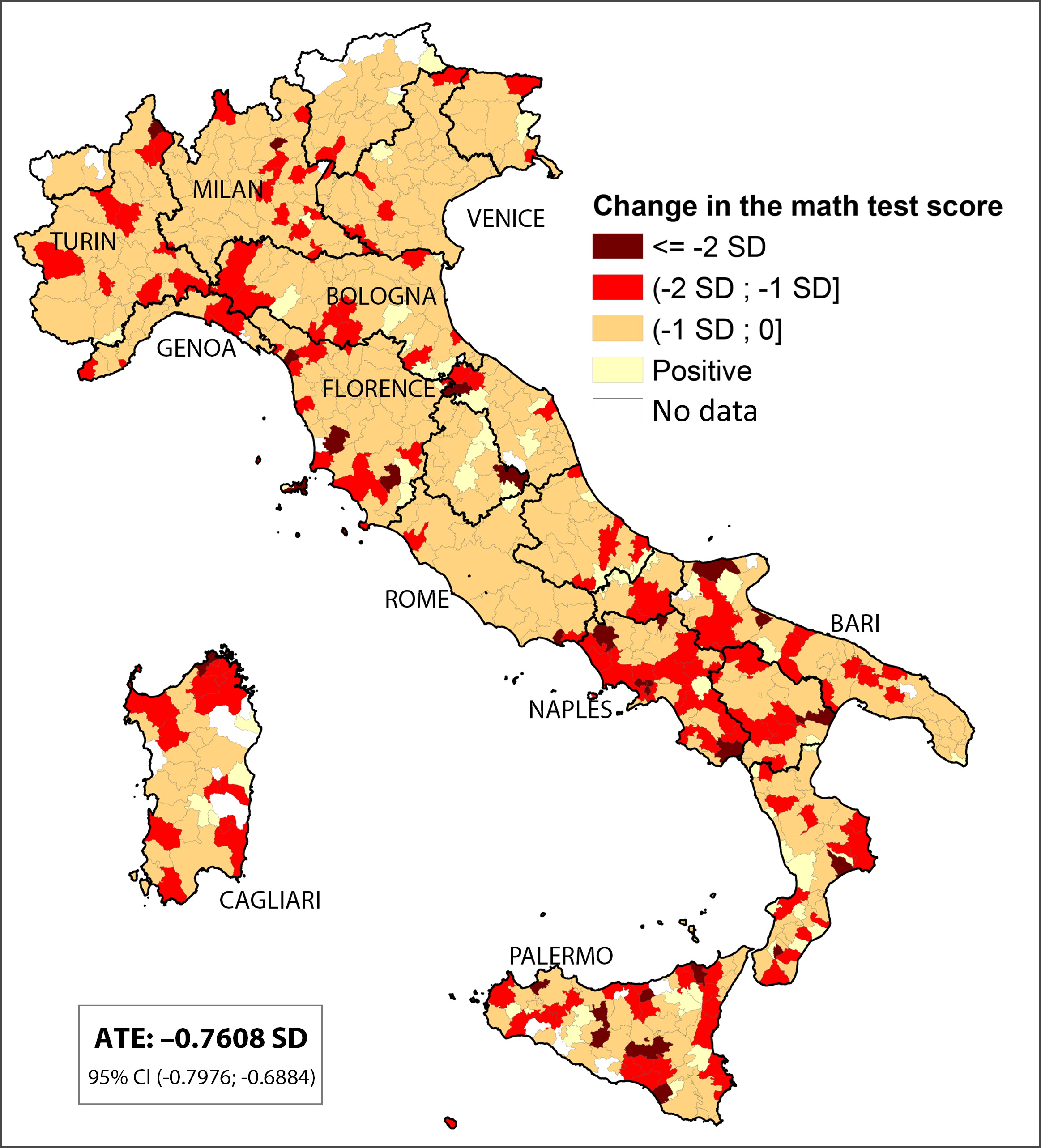}
    \caption{The local impact of the COVID-19 pandemic on education.}
    \label{fig:ate}
\end{figure}

Figure \ref{fig:ate} shows the local effects of the pandemic on math scores of fifth-grade students across Italian LLMs and reports the ATE. These estimates come from the best-performing technique of the Design Stage, the MLCM with random forest.\footnote{Figure \ref{appfig:shapley} in Supplemental Appendix \ref{app:application} provides SHAP values for this model for the counterfactual 2020 forecasts.} 

The first insight is that the COVID-19 shock engendered major learning losses in Italy. The ATE points to a decrease in the math score of –0.7608 standard deviation (SD) with respect to the pre-pandemic average, and this effect is strongly statistically significant (95\% confidence intervals: –0.7976; –0.6884). This result is in line with micro-level studies documenting large negative impacts on educational outcomes in Italy \citep{Battisti:Maggio:2023, Carlana:Ferrara:Lopez:2023}. 

As shown above, some of the smallest LLMs (the most volatile) were imprecisely estimated in the placebo exercise. Since equal weight is attributed to all treated units, the presence of these LLMs might affect the accuracy of the ATE estimate. To address this potential concern, we recommend a sensitivity analysis that re-estimates the ATE following the exclusion of units with the poorest fit in the pre-treatment period. Here, we removed the top 1\%, 2\%, and 5\% of the most imprecisely estimated LLMs in the placebo tests. The resultant ATE estimates were -0.7500, -0.7568, and -0.7341, respectively—each closely aligning with the ATE estimate of -0.7608.

The second key finding is that this average impact masks considerable heterogeneity across the Italian territory. Some LLMs experienced much larger drops in students’ performances. In particular, the MLCM detects learning losses of more than one standard deviation in clusters of local economies mostly located in the South of Italy (Campania, Apulia, Molise, Calabria, Sicily, and Sardinia), and drops in math scores greater than two SD in scattered areas predominantly concentrated in Central and Southern regions.

\begin{figure}
    \centering
    \includegraphics[scale = 0.9]{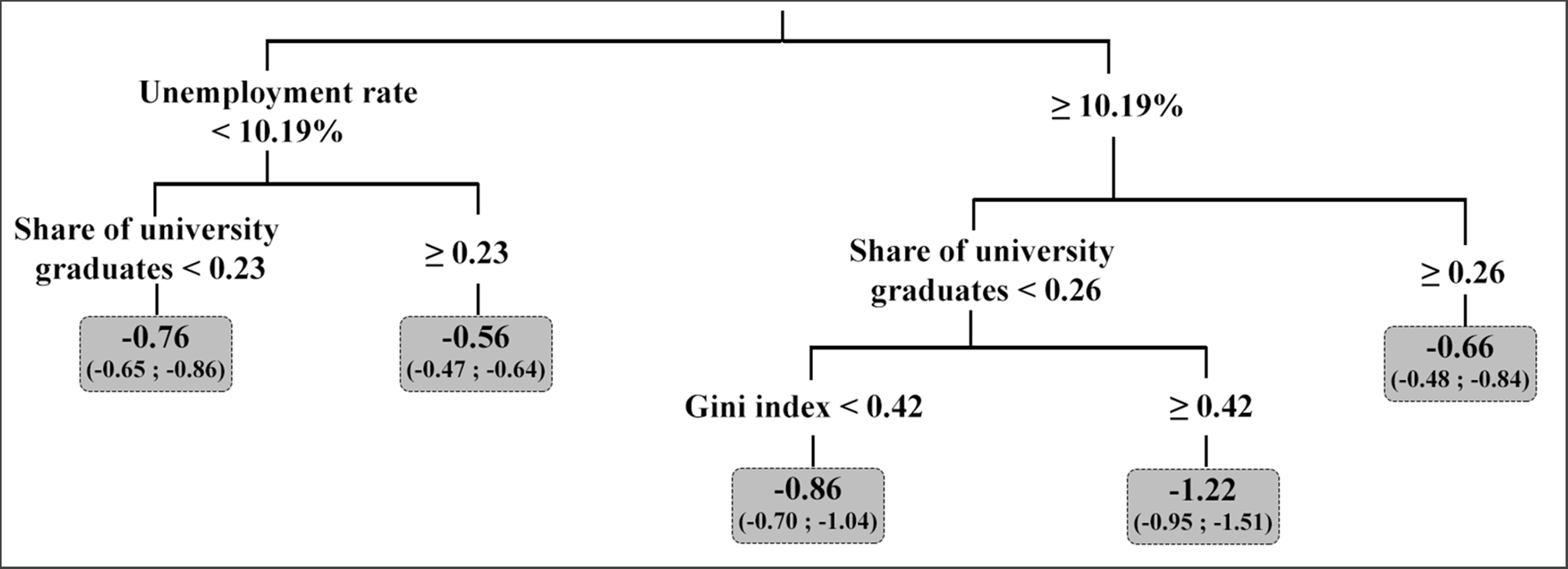}
    \caption{Data-driven CATEs. \textit{Notes}: The values within the terminal nodes report the CATEs (measured in SD) for all LLMs grouped within that node. Lower and upper bounds  refer to the 95\% confidence interval and are reported in parentheses.}
    \label{fig:cate}
\end{figure}

Figure \ref{fig:cate} presents data-driven CATEs estimated with the regression tree analysis.\footnote{For this analysis, we selected a set of LLM-level variables including, among others, pre-pandemic variables such as the per capita income, the unemployment rate, the Gini index, the share of university graduates, two proxies of access to distance learning that capture the digital divide across areas, and excess deaths registered during the first wave of the pandemic \citep{Cerqua:DiStefano:Letta:Miccoli:2021}. The full list can be found in Table \ref{tabapp:cate_vars} in Supplemental Appendix \ref{app:application}.} The algorithm selects only three predictors to construct the tree. The analysis reveals that the most substantial learning losses (1.22 SD below the pre-COVID baseline) occurred in LLMs characterized by an unemployment rate equal to or above 10.19\%, a percentage of university graduates below 26\%, and a Gini index equal to or above 0.42. Conversely, the territories with an above-average treatment effect (–0.56 SD) exhibit an unemployment rate below 10.19\% and a percentage of university graduates equal to or above 23\%. Finally, all the CATEs reported in the tree are statistically significant.

Since areas with higher unemployment, greater inequality, and lower education rates at baseline have experienced disproportionate impacts, the pre-existing gaps across the country have been amplified. If such gaps persist, it is likely that they will act as a catalyst for future economic inequality, leading to widened territorial disparities in the medium and long term. In summary, we can anticipate that, in the absence of counterbalancing policy efforts, the pandemic will exacerbate long-standing inequalities that predate COVID-19.

\section{Conclusions}
\label{sec:conclusion}
Identifying causal effects is challenging without a credible control group. We overcome this challenge by proposing a new method based on counterfactual forecasting with machine learning. The MLCM can employ any off-the-shelf ML algorithm to estimate policy-relevant causal parameters in a wide variety of panel settings, including very short panels, without relying on untreated units. To illustrate its potential, we presented numerical studies, a replication of the minimum wage application in \citet{callaway2021difference} without using untreated units, and an empirical analysis on the effects of the COVID-19 crisis on education in Italy. The companion \textsf{R} package \href{https://github.com/FMenchetti/MachineControl}{\textcolor{blue}{\texttt{MachineControl}}} provides an easy-to-use implementation of the proposed approach. Large shocks, international economic policies, nationwide policy changes, and regional programs engendering substantial interaction between units, are all real-world scenarios where a control group may not exist. In such cases, researchers can harness the MLCM, which complements the existing econometric toolbox for causal inference and policy evaluation.


\newpage
\bibliographystyle{chicago}
\bibliography{references}  

\newpage
\renewcommand{\appendixpagename}{Supplemental Appendix}
\begin{appendices}
\renewcommand{\appendixpagename}{Supplemental Appendix}

\section{Causal estimands in settings with interference between units}
\counterwithin{table}{section}  
\counterwithin{figure}{section} 

\label{app:additional_estimands}
Compared to Case (i), in the scenario with violations of the no-interference assumption, the causal estimands are different. In this section, we define the main causal estimands for the setup in which only a subgroup of units is treated, but there are spillover and general equilibrium effects that prevent using the set of untreated units to build a valid control group (Case (ii) defined above). 
We can first define the ATT across the treated units at a given point in time.
\begin{defi}
For any strictly positive integer $k$, let $\tau_{i,t_0+k} = \Y_{i,t_0+k}(1) - \Y_{i,t_0+k|t_0}(0)$ be the unit-level treatment effect of the policy at time $t_0+k$ and denote with $N_T$ the number of treated units. The ATT at time $t_0+k$ is defined as,
\begin{equation*}
    \tau_{t_0+k}^T = \frac{1}{N_T} \sum_{i = 1}^{N_T} \tau_{i,t_0+k} = \frac{1}{N_T} \sum_{i = 1}^{N_T} \Y_{i,t_0+k}(1) - \Y_{i,t_0+k|t_0}(0).
\end{equation*}
\end{defi}
Note that, in this framework, the ATE does not coincide with the ATT because not all units are directly exposed to the treatment.

As a subgroup of the units is not treated but potentially subject to spillovers, we can define the Average Spillover effect on the Affected (ASA) across the untreated units at a given point in time.
\begin{defi}
    For any strictly positive integer $k$, let $\tau_{j,t_0+k} = \Y_{j,t_0+k}(1) - \Y_{j,t_0+k|t_0}(0)$ denote the unit-level spillover effect of the policy at time $t_0+k$ for all untreated units and denote with $N_C$ the number of untreated units. The ASA at time $t_0+k$ is defined as,
    \begin{equation*}
         \tau_{t_0+k}^C = \frac{1}{N_C} \sum_{i = 1}^{N_C} \tau_{j,t_0+k} = \frac{1}{N_C} \sum_{i = 1}^{N_C} \Y_{j,t_0+k}(1) - \Y_{j,t_0+k|t_0}(0).
    \end{equation*}
\end{defi}
Both causal estimands, $\tau_{t_0+k}^T$ and $\tau_{t_0+k}^C$, can be estimated with the MLCM by applying the same implementation routine defined in the paper separately for the group of treated and untreated units.

\newpage

\section{Identification proof}
\label{app:proof_general}
We show here the general identification proof for the causal estimands defined in Section \ref{subsec:estimands}.

\begin{proof}
As in the main text, the proof is organized by distinguishing between $k = 1$ and $k = 2$.
\begin{itemize}
    \item $k = 1$. Let $\Y_{i,t}(0)$ follow Equation (\ref{eqn:model_assumpt}). In the expression $\tau_{i,t_0+1} = \Y_{i,t_0+1}(1) - \Y_{i,t_0+1|t_0}(0)$, the first term $\Y_{i,t_0+1}(1)$ is the observed outcome under the policy and thus is immediately identified. Since $Q = \max{ \{ 0, q \} } = q$, the second term can be written as 
    \begin{small}
    \begin{align*} 
        \Y_{i,t_0+1|t_0}(0) & = \E[\Y_{i,t_0+1}(0)|\Y_{i,t_0}^{(p)}, \bb{X}_{i,t_0+1}^{(q)}] & \text{Def. of $\Y_{i,t_0+1|t_0}(0)$} \\ 
        & = \E[h(\Y_{i,t_0}^{(p)}(0), \bb{X}_{i,t_0+1}^{(q)}) |\Y_{i,t_0}^{(p)}, \bb{X}_{i,t_0+1}^{(q)}]  & \text{A.\ref{assumpt:model} - Eq.(\ref{eqn:model_assumpt})} \\       
        & = h(y_{i,t_0}^{(p)}(0), \bb{x}_{i,t_0+1}^{(q)}) = y_{i,t_0+1|t_0}(0).
    \end{align*}
    \end{small}
    The last expression follows from the fact that $h(\cdot)$ is deterministic once we observe $\Y_{i,t_0}^{(p)}(0) = y_{i,t_0}^{(p)}(0)$ and $\bb{X}^{(q)}_{i,t_0+1} = \bb{x}^{(q)}_{i,t_0+1}$. As a result, $\Y_{i,t_0+1|t_0}(0)$ is identified from observed data, and so is $\tau_{i,t_0+1}$. Notice that in this proof, we never used the linearity of the potential outcome model. Thus, the proof at $k = 1$ is the same even when $h(\cdot)$ is non-linear.  
   
    \item $k = 2$. In the expression $\tau_{i,t_0+2} = \Y_{i,t_0+2}(1) - \Y_{i,t_0+2|t_0}(0)$, the first term $\Y_{i,t_0+2}(1)$ is the observed outcome under the policy. Thus, the identification of the causal effect depends solely on the second term, $\Y_{i,t_0+2|t_0}(0)$. Since $Q = \max{ \{ 1, q\} }$, we now have two possible cases: either $q \geq 1$, simplifying to $Q = q$ or $q = 0$, in which case $Q = 1 > q$. Assuming the latter, we now distinguish between a linear and a non-linear model specification:
    \end{itemize}
    \begin{itemize}    
        \item[-] Linear case. Assume the following linear model $h(\Y_{i,t-1}^{(p)}(0), \bb{X}_t^{(q)}) = \sum_{j = 0}^p b_1^{(j)}\Y_{i,t-1-j}(0) + \sum_{j = 0}^q \bb{b_2} \bb{X}_{i,t-q}$. The term $\Y_{i,t_0+2|t_0}(0)$ can then be written as,
        \begin{small}
        \begin{align*}
         \Y_{i,t_0+2|t_0}(0)  & = \E[\Y_{i,t_0+2}(0)|\Y_{i,t_0}^{(p)}, \bb{X}_{i,t_0+2}, \bb{X}_{i,t_0+1}] \hspace{143pt} \text{Def. $\Y_{i,t_0+2|t_0}(0)$}  \\         
        & = \E[h(\Y_{i,t_0+1}^{(p)}(0), \bb{X}_{i,t_0+2})|\Y_{i,t_0}^{(p)}, \bb{X}_{i,t_0+2}, \bb{X}_{i,t_0+1} ] \hspace{113pt} \text{A.\ref{assumpt:model} - Eq.(\ref{eqn:model_assumpt})} \\
        & = \E [b_1^{(0)} \Y_{i,t_0+1}(0) + \sum_{j = 1}^p  b_1^{(j)} \Y_{i,t_0+1-j} + \bb{b_2} \bb{X}_{i,t_0+2}|\Y_{i,t_0}^{(p)}, \bb{X}_{i,t_0+2}, \bb{X}_{i,t_0+1}] \hspace{15pt} \text{Linearity } \\
        & = b_1^{(0)} \E[\Y_{i,t_0+1}(0)|\Y_{i,t_0}^{(p)}, \bb{X}_{i,t_0+1}] + \sum_{j = 1}^p  b_1^{(j)} y_{i,t_0+1-j} + \bb{b_2} \bb{x}_{i,t_0+2} \hspace{83pt} \text{A.\ref{assumpt:no_anticipation}} \\
        & = b_1^{(0)} y_{i,t_0+1|t_0}(0) + \sum_{j = 1}^p  b_1^{(j)} y_{i,t_0+1-j} + \bb{b_2} \bb{x}_{i,t_0+2}.      
    \end{align*}
    \end{small}
    \item[-] Non-linear case. The identification of $\Y_{i,t_0+2|t_0}(0)$ can be based on the formulation for the post-intervention potential outcome model defined by Equation (\ref{eqn:direct_models}),
    \begin{small}    
    \begin{align*}
         \Y_{i,t_0+2|t_0}(0)  & = \E[\Y_{i,t_0+2}(0)|\Y_{i,t_0}(0), \bb{X}_{i,t_0+2}, \bb{X}_{i,t_0+1}] \hspace{105pt} \text{Def. $\Y_{i,t_0+2|t_0}(0)$} \\         
        & = \E[g_2(\Y_{i,t_0+1|t_0}(0),\Y_{i,t_0}^{(p)}, \bb{X}_{i,t_0+2})|\Y_{i,t_0}^{(p)}, \bb{X}_{i,t_0+2}, \bb{X}_{i,t_0+1}] \hspace{50pt} \text{A.\ref{assumpt:additional_models} - Eq.(\ref{eqn:direct_models})}  \\
        & = \E[g_2(h(\Y_{i,t_0}^{(p)}, \bb{X}_{i,t_0+1}),\Y_{i,t_0}^{(p)}, \bb{X}_{i,t_0+2})|\Y_{i,t_0}^{(p)}, \bb{X}_{i,t_0+2}, \bb{X}_{i,t_0+1}]  \\
        & =g_2(y_{i,t_0+1|t_0}(0), y_{i,t_0}^{(p)},\bb{x}_{i,t_0+2}).
    \end{align*}
    \end{small}
    The last expression follows from the fact that $g_2(\cdot)$ is deterministic once we observe $\Y_{i,t_0} = y_{i,t_0}$, $\bb{X}_{i,t_0+1} = \bb{x}_{i,t_0+1}$ and  $\bb{X}_{i,t_0+2} = \bb{x}_{i,t_0+2}$. In addition, even though Assumption \ref{assumpt:no_anticipation} was not explicitly mentioned in this part of the proof, notice that it is implicit in Assumption \ref{assumpt:additional_models}, as by Equation (\ref{eqn:direct_models}), the potential outcomes never depend on future covariates.
    \end{itemize}
     
The proof for a generic time $t_0+k$ follows analogously by induction. 
\end{proof}

\clearpage

\newpage
\newpage
\section{Bootstrap inference for ATE and CATE confidence intervals}
\label{app:bootstrap}

Following recent work on causal machine learning estimators for panel data \citep{Viviano:Bradic:2023}, inference on causal effects estimated with the MLCM is performed by block-bootstrap.
Since we are dealing with a panel dataset that has an inherent temporal structure, resampling the unit-time pairs independently in a context with dependent data would lead to data leakage. Therefore, we sample the units instead of the unit-time pairs so that if a unit is sampled, its entire evolution over time is sampled as well. This way, the inference is more robust to misspecification issues. This idea is in the spirit of the original block-bootstrap introduced by \citet{Carlstein:1986} and \citet{Kunsch:1989} that consists in dividing a time series in blocks (overlapping or non-overlapping) and then sampling the blocks with replacement. Since we have a panel dataset, we can consider each unit to form one block. The full algorithm to derive block-bootstrap confidence intervals in our setting is detailed in Algorithm \ref{algorithm_ATE} below.

\begin{algorithm}[h!]
\begin{algorithmic}[1]
\caption{Block-bootstrap algorithm for the ATE}
\label{algorithm_ATE}
\REQUIRE $N$, $B$, 
\FOR{b in $1:B$ where $B$ is the total number of bootstrap iterations} 
\STATE Sample $N$ units with replacement
\STATE Only for the resampled units, retain the tuple $({\Y}_{i,t}^{(p)}(0),\bb{X}_{i,t}^{(q)})^{(b)}$ for all $t = 1, \dots, T$ (pre- and post-intervention)
\STATE Compute the bootstrap replication of the ATE, i.e., $    \hat{\tau}^{(b)}_t$ by repeating the Design stage (steps 3 to 5) and the Analysis stage (steps 6 to 9) of the MLCM outlined in Section \ref{sec:implementation}
\ENDFOR
\STATE Let  $\widehat{G}(c) = \# \{\hat{\tau}^{(b)}_t \leq c \}/B$ be the empirical cumulative distribution function of the $B$ bootstrap replications. Compute a $(1- \alpha)$  percentile interval as $[\widehat{G}^{-1} (\alpha/2), \widehat{G}^{-1}(1-\alpha/2) ]$.
\end{algorithmic}
\end{algorithm}

Deriving confidence intervals for the CATEs is more challenging than for the ATE. The reason is that CATEs are estimated with a regression tree where the outcome is a collection of individual causal effects. Therefore, if we resample the units at the very beginning and then we estimate $B$ different CATEs, we will end up with very different trees: both the splits and the observations within each terminal node will likely differ. For this reason, we propose to resample directly the observations within each terminal node of the tree. Note that the observations are the estimated individual causal effects (computed by comparing the observed data with the ML forecasts). Our estimand of interest is the group-average of the individual effects, so at each bootstrap iteration, we resample the units in each terminal node, averaging the individual effects and obtaining a bootstrap distribution for the CATE. The full algorithm to derive confidence intervals is described in Algorithm \ref{algorithm_CATE} below.

\begin{algorithm}[H]
\begin{algorithmic}[1]
\caption{Block-bootstrap algorithm for CATEs}
\label{algorithm_CATE}
\REQUIRE $\hat{\tau}_t$, $\bb{X}_{i,t}$, $D$, $B$, 
\FOR{b in $1:B$ where $B$ is the total number of bootstrap iterations and for $d = 1, \dots,D$ where $D$ is the total number of terminal nodes of the tree} 
\STATE Sample with replacement the $N_d$ units in terminal node $d$
\STATE Only for the resampled units, retain the tuple $(\hat{\tau}_{i,t}, \bb{X}_{i,t})_d^{(b)}$ 
\STATE Compute the bootstrap replication of CATE in the terminal node $d$ by averaging the ATEs of the resampled units, i.e., $\hat{\tau}_t(d)$. 
\STATE By definition (\ref{eqn:hat_ate_cate})$,    \hat{\tau}_t(d)$ is an estimator of CATE, since we are averaging the individual effects among the units in group (terminal node) $d$, i.e., we are conditioning on covariates
\ENDFOR
\STATE Let  $\widehat{G}(c) = \# \{\hat{\tau}_t(d) \leq c \}/B$ be the empirical cumulative distribution function of the $d$-group CATE in the $B$ bootstrap replication. Compute a $(1- \alpha)$  percentile interval as $[\widehat{G}^{-1} (\alpha/2), \widehat{G}^{-1}(1-\alpha/2) ]$.
\end{algorithmic}
\end{algorithm}
\newpage
\section{Additional simulation results}
\label{app:simulation}

In this Section of the Supplemental Appendix, we report the results of all the simulation scenarios that we generated by combining the following parameters: i) number of units in the panel ($N = 400$, $N = 200$); ii) autoregressive coefficient ($\phi = 0.8$, $\phi = 1.2)$; iii) standard deviation of $u_i$ regulating the amount of heterogeneity between units ($\sigma_u = 1$, $\sigma_u = 0.1$).

Table \ref{tab:n400phi12u1} shows the results of an ``explosive'' simulation scenario corresponding to the one reported in the main text ($N = 400$, $\sigma_u = 1$), but with $\phi = 1.2$. Tables \ref{tab:n400phi08u01} and \ref{tab:n400phi12u01} report the results for the remaining combinations with $N = 400$ and $\sigma_u = 0.1$.

Tables \ref{tab:n200phi08u1}--\ref{tab:n200phi12u01} show the results of the simulations when the number of units in the panel reduces to $N = 200$.
\begin{table}[h] 
\caption{Simulation results for $N = 400$, $\phi = 1.2$ and $\sigma_u = 1$}
\label{tab:n400phi12u1}
\scalebox{0.85}{
\begin{tabular}{cccccccccccc} \hline 
\multicolumn{12}{c}{1st Post-intervention period ($t_0+1$)}  \\ \hline
& & & \multicolumn{4}{c}{Linear} & & \multicolumn{4}{c}{Non-Linear} \\ \cline{4-7} \cline{9-12}
Pre-int.          & Bootstrap   &  & True ATE  & Bias  & Rel. Bias  & Coverage     &      & True ATE  & Bias  & Rel. Bias  & Coverage \\
4 & block & & 234.28 & 0.17 & 0.001 & 0.94 & & 4.07 & 0.10 & 0.024 & 0.95 \\ 
9 & block & & 641.62 & 0.12 & 0.000 & 0.94 & & 4.14 & 0.09 & 0.023 & 0.95 \\ \hline
\multicolumn{12}{c}{2nd Post-intervention period ($t_0+2$)}  \\ \hline
& & & \multicolumn{4}{c}{Linear} & & \multicolumn{4}{c}{Non-Linear} \\ \cline{4-7} \cline{9-12}
Pre-int.          & Bootstrap   &  & True ATE  & Bias  & Rel. Bias  & Coverage     &      & True ATE  & Bias  & Rel. Bias  & Coverage \\
4 & block & & 175.71 & 0.39 & 0.002 & 0.93 & & 3.05 & 0.10 & 0.032 & 0.91 \\ 
9 & block & & 481.21 & 0.25 & 0.001 & 0.92 & & 3.10 & 0.10 & 0.031 & 0.92 \\ \hline
\multicolumn{12}{c}{3rd Post-intervention period ($t_0+3$)}  \\ \hline
& & & \multicolumn{4}{c}{Linear} & & \multicolumn{4}{c}{Non-Linear} \\ \cline{4-7} \cline{9-12}
Pre-int.           & Bootstrap   &  & True ATE  & Bias  & Rel. Bias  & Coverage     &      & True ATE  & Bias  & Rel. Bias  & Coverage \\
4 & block & & 117.14 & 0.75 & 0.006 & 0.93 & & 2.03 & 0.10 & 0.050 & 0.92 \\ 
9 & block & & 320.81 & 0.46 & 0.001 & 0.94 & & 2.07 & 0.09 & 0.045 & 0.92 \\ 
   \hline
\end{tabular}}
\end{table}

\begin{table}[H]
\centering
\caption{Simulation results for $N = 400$, $\phi = 0.8$ and $\sigma_u = 0.1$}
\label{tab:n400phi08u01}
\scalebox{0.85}{
\begin{tabular}{cccccccccccc} \hline 
\multicolumn{12}{c}{1st Post-intervention period ($t_0+1$)}  \\ \hline
& & & \multicolumn{4}{c}{Linear} & & \multicolumn{4}{c}{Non-Linear} \\ \cline{4-7} \cline{9-12}
Pre-int.           & Bootstrap   &  & True ATE  & Bias  & Rel. Bias  & Coverage     &      & True ATE  & Bias  & Rel. Bias  & Coverage \\
4 & block & & 73.38 & 0.27 & 0.004 & 0.97 & & 4.07 & 0.14 & 0.033 & 0.97 \\ 
9 & block & & 92.58 & 0.15 & 0.002 & 0.95 & & 4.14 & 0.11 & 0.027 & 0.95 \\ \hline 
\multicolumn{12}{c}{2nd Post-intervention period ($t_0+2$)}  \\ \hline
& & & \multicolumn{4}{c}{Linear} & & \multicolumn{4}{c}{Non-Linear} \\ \cline{4-7} \cline{9-12}
Pre-int.           & Bootstrap   &  & True ATE  & Bias  & Rel. Bias  & Coverage     &      & True ATE  & Bias  & Rel. Bias  & Coverage \\
4 & block & & 55.03 & 0.37 & 0.007 & 0.94 & & 3.05 & 0.10 & 0.034 & 0.91 \\ 
9 & block & & 69.44 & 0.21 & 0.003 & 0.93 & & 3.10 & 0.11 & 0.034 & 0.93 \\ \hline
\multicolumn{12}{c}{3rd Post-intervention period ($t_0+3$)}  \\ \hline
& & & \multicolumn{4}{c}{Linear} & & \multicolumn{4}{c}{Non-Linear} \\ \cline{4-7} \cline{9-12}
Pre-int.           & Bootstrap   &  & True ATE  & Bias  & Rel. Bias  & Coverage     &      & True ATE  & Bias  & Rel. Bias  & Coverage \\
4 & block & & 36.69 & 0.72 & 0.020 & 0.94 & & 2.04 & 0.14 & 0.071 & 0.95 \\ 
9 & block & & 46.29 & 0.33 & 0.007 & 0.94 & & 2.07 & 0.11 & 0.051 & 0.93 \\ 
   \hline
\end{tabular}}
\end{table}

\begin{table}[H]
\centering
\caption{Simulation results for $N = 400$, $\phi = 1.2$ and $\sigma_u = 0.1$}
\label{tab:n400phi12u01}
\scalebox{0.85}{
\begin{tabular}{cccccccccccc} \hline 
\multicolumn{12}{c}{1st Post-intervention period ($t_0+1$)}  \\ \hline
& & & \multicolumn{4}{c}{Linear} & & \multicolumn{4}{c}{Non-Linear} \\ \cline{4-7} \cline{9-12}
Pre-int.           & Bootstrap   &  & True ATE  & Bias  & Rel. Bias  & Coverage     &      & True ATE  & Bias  & Rel. Bias  & Coverage \\ 
4 & block & & 233.97 & 0.31 & 0.001 & 0.96 & & 4.07 & 0.14 & 0.034 & 0.97 \\ 
9 & block & & 640.02 & 0.18 & 0.000 & 0.95 & & 4.14 & 0.11 & 0.027 & 0.95 \\ \hline
\multicolumn{12}{c}{2nd Post-intervention period ($t_0+2$)}  \\ \hline
& & & \multicolumn{4}{c}{Linear} & & \multicolumn{4}{c}{Non-Linear} \\ \cline{4-7} \cline{9-12}
Pre-int.           & Bootstrap   &  & True ATE  & Bias  & Rel. Bias  & Coverage     &      & True ATE  & Bias  & Rel. Bias  & Coverage \\
4 & block & & 175.47 & 0.65 & 0.004 & 0.95 & & 3.05 & 0.10 & 0.034 & 0.92 \\ 
9 & block & & 480.01 & 0.36 & 0.001 & 0.94 & & 3.11 & 0.11 & 0.034 & 0.94 \\ \hline
\multicolumn{12}{c}{3rd Post-intervention period ($t_0+3$)}  \\ \hline
& & & \multicolumn{4}{c}{Linear} & & \multicolumn{4}{c}{Non-Linear} \\ \cline{4-7} \cline{9-12}
Pre-int.           & Bootstrap   &  & True ATE  & Bias  & Rel. Bias  & Coverage     &      & True ATE  & Bias  & Rel. Bias  & Coverage \\
4 & block & & 116.98 & 1.34 & 0.011 & 0.95 & & 2.03 & 0.15 & 0.073 & 0.95 \\ 
9 & block & & 320.01 & 0.72 & 0.002 & 0.95 & & 2.07 & 0.10 & 0.051 & 0.94 \\ 
   \hline
\end{tabular}}
\end{table}

\begin{table}[H]
\centering
\caption{Simulation results for $N = 200$ units, $\phi = 0.8$ and $\sigma_u = 1$}
\label{tab:n200phi08u1}
\scalebox{0.85}{
\begin{tabular}{cccccccccccc} \hline 
\multicolumn{12}{c}{1st Post-intervention period ($t_0+1$)}  \\ \hline
& & & \multicolumn{4}{c}{Linear} & & \multicolumn{4}{c}{Non-Linear} \\ \cline{4-7} \cline{9-12}
Pre-int.           & Bootstrap   &  & True ATE  & Bias  & Rel. Bias  & Coverage     &      & True ATE  & Bias  & Rel. Bias  & Coverage \\ 
4 & block & & 74.23 & 0.22 & 0.003 & 0.93 & & 4.07 & 0.13 & 0.033 & 0.97 \\ 
9 & block & & 92.07 & 0.15 & 0.002 & 0.94 & & 4.14 & 0.13 & 0.032 & 0.94 \\ \hline
\multicolumn{12}{c}{2nd Post-intervention period ($t_0+2$)}  \\ \hline
& & & \multicolumn{4}{c}{Linear} & & \multicolumn{4}{c}{Non-Linear} \\ \cline{4-7} \cline{9-12}
Pre-int.           & Bootstrap   &  & True ATE  & Bias  & Rel. Bias  & Coverage     &      & True ATE  & Bias  & Rel. Bias  & Coverage \\ 
4 & block & & 55.67 & 0.36 & 0.006 & 0.90 & & 3.05 & 0.13 & 0.044 & 0.92 \\ 
9 & block & & 69.06 & 0.22 & 0.003 & 0.90 & & 3.11 & 0.13 & 0.042 & 0.93 \\ \hline
\multicolumn{12}{c}{3rd Post-intervention period ($t_0+3$)}  \\ \hline
& & & \multicolumn{4}{c}{Linear} & & \multicolumn{4}{c}{Non-Linear} \\ \cline{4-7} \cline{9-12}
Pre-int.           & Bootstrap   &  & True ATE  & Bias  & Rel. Bias  & Coverage     &      & True ATE  & Bias  & Rel. Bias  & Coverage \\ 
4 & block & & 37.11 & 0.48 & 0.013 & 0.93 & & 2.04 & 0.14 & 0.068 & 0.93 \\ 
9 & block & & 46.04 & 0.29 & 0.006 & 0.92 & & 2.07 & 0.13 & 0.064 & 0.92 \\ 
   \hline
\end{tabular}}
\end{table}

\begin{table}[H]
\centering
\caption{Simulation results for $N = 200$ units, $\phi = 0.8$ and $\sigma_u = 0.1$}
\label{tab:n200phi08u01}
\scalebox{0.85}{
\begin{tabular}{cccccccccccc} \hline 
\multicolumn{12}{c}{1st Post-intervention period ($t_0+1$)}  \\ \hline
& & & \multicolumn{4}{c}{Linear} & & \multicolumn{4}{c}{Non-Linear} \\ \cline{4-7} \cline{9-12}
Pre-int.           & Bootstrap   &  & True ATE  & Bias  & Rel. Bias  & Coverage     &      & True ATE  & Bias  & Rel. Bias  & Coverage \\ 
4 & block & & 73.34 & 0.41 & 0.006 & 0.96 & & 4.07 & 0.19 & 0.046 & 0.96 \\ 
9 & block & & 91.86 & 0.21 & 0.002 & 0.94 & & 4.14 & 0.15 & 0.037 & 0.95 \\ \hline
\multicolumn{12}{c}{2nd Post-intervention period ($t_0+2$)}  \\ \hline
& & & \multicolumn{4}{c}{Linear} & & \multicolumn{4}{c}{Non-Linear} \\ \cline{4-7} \cline{9-12}
Pre-int.           & Bootstrap   &  & True ATE  & Bias  & Rel. Bias  & Coverage     &      & True ATE  & Bias  & Rel. Bias  & Coverage \\
4 & block & & 55.01 & 0.56 & 0.010 & 0.93 & & 3.05 & 0.15 & 0.048 & 0.91 \\ 
9 & block & & 68.90 & 0.32 & 0.005 & 0.92 & & 3.11 & 0.15 & 0.048 & 0.94 \\ \hline
\multicolumn{12}{c}{3rd Post-intervention period ($t_0+3$)}  \\ \hline
& & & \multicolumn{4}{c}{Linear} & & \multicolumn{4}{c}{Non-Linear} \\ \cline{4-7} \cline{9-12}
Pre-int.           & Bootstrap   &  & True ATE  & Bias  & Rel. Bias  & Coverage     &      & True ATE  & Bias  & Rel. Bias  & Coverage \\
4 & block & & 36.67 & 1.07 & 0.029 & 0.92 & & 2.03 & 0.22 & 0.110 & 0.95 \\ 
9 & block & & 45.93 & 0.50 & 0.011 & 0.94 & & 2.07 & 0.15 & 0.074 & 0.93 \\ 
   \hline
\end{tabular}}
\end{table}

\begin{table}[H]
\centering
\caption{Simulation results for $N = 200$ units, $\phi = 1.2$ and $\sigma_u = 1$}
\label{tab:n200phi12u1}
\scalebox{0.85}{
\begin{tabular}{cccccccccccc} \hline 
\multicolumn{12}{c}{1st Post-intervention period ($t_0+1$)}  \\ \hline
& & & \multicolumn{4}{c}{Linear} & & \multicolumn{4}{c}{Non-Linear} \\ \cline{4-7} \cline{9-12}
Pre-int.           & Bootstrap   &  & True ATE  & Bias  & Rel. Bias  & Coverage     &      & True ATE  & Bias  & Rel. Bias  & Coverage \\ 
4 & block & & 232.58 & 0.26 & 0.001 & 0.94 & & 4.07 & 0.14 & 0.033 & 0.96 \\ 
9 & block & & 629.30 & 0.18 & 0.000 & 0.94 & & 4.14 & 0.14 & 0.033 & 0.94 \\ \hline
\multicolumn{12}{c}{2nd Post-intervention period ($t_0+2$)}  \\ \hline
& & & \multicolumn{4}{c}{Linear} & & \multicolumn{4}{c}{Non-Linear} \\ \cline{4-7} \cline{9-12}
Pre-int.           & Bootstrap   &  & True ATE  & Bias  & Rel. Bias  & Coverage     &      & True ATE  & Bias  & Rel. Bias  & Coverage \\ 
4 & block & & 174.44 & 0.62 & 0.004 & 0.93 & & 3.05 & 0.14 & 0.045 & 0.93 \\ 
9 & block & & 471.97 & 0.37 & 0.001 & 0.92 & & 3.11 & 0.13 & 0.042 & 0.93 \\ \hline
\multicolumn{12}{c}{3rd Post-intervention period ($t_0+3$)}  \\ \hline
& & & \multicolumn{4}{c}{Linear} & & \multicolumn{4}{c}{Non-Linear} \\ \cline{4-7} \cline{9-12}
Pre-int.           & Bootstrap   &  & True ATE  & Bias  & Rel. Bias  & Coverage     &      & True ATE  & Bias  & Rel. Bias  & Coverage \\ 
4 & block & & 116.29 & 1.18 & 0.010 & 0.93 & & 2.03 & 0.14 & 0.069 & 0.93 \\ 
9 & block & & 314.65 & 0.67 & 0.002 & 0.93 & & 2.07 & 0.13 & 0.064 & 0.91 \\ 
   \hline
\end{tabular}}
\end{table}

\begin{table}[H]
\centering
\caption{Simulation results for $N = 200$ units, $\phi = 1.2$ and $\sigma_u = 0.1$}
\label{tab:n200phi12u01}
\scalebox{0.85}{
\begin{tabular}{cccccccccccc} \hline 
\multicolumn{12}{c}{1st Post-intervention period ($t_0+1$)}  \\ \hline
& & & \multicolumn{4}{c}{Linear} & & \multicolumn{4}{c}{Non-Linear} \\ \cline{4-7} \cline{9-12}
Pre-int.           & Bootstrap   &  & True ATE  & Bias  & Rel. Bias  & Coverage     &      & True ATE  & Bias  & Rel. Bias  & Coverage \\ 
4 & block & & 74.24  & 0.14 & 0.002 & 0.94 & & 4.07 & 0.19 & 0.046 & 0.96 \\ 
9 & block & & 632.26 & 0.27 & 0.000 & 0.94 & & 4.14 & 0.15 & 0.037 & 0.96 \\ \hline
\multicolumn{12}{c}{2nd Post-intervention period ($t_0+2$)}  \\ \hline
& & & \multicolumn{4}{c}{Linear} & & \multicolumn{4}{c}{Non-Linear} \\ \cline{4-7} \cline{9-12}
Pre-int.           & Bootstrap   &  & True ATE  & Bias  & Rel. Bias  & Coverage     &      & True ATE  & Bias  & Rel. Bias  & Coverage \\ 
4 & block & & 55.68  & 0.22 & 0.004 & 0.92 & & 3.050 & 0.14 & 0.047 & 0.91 \\ 
9 & block & & 474.20 & 0.55 & 0.001 & 0.93 & & 3.100 & 0.14 & 0.046 & 0.95 \\ \hline
\multicolumn{12}{c}{3rd Post-intervention period ($t_0+3$)}  \\ \hline
& & & \multicolumn{4}{c}{Linear} & & \multicolumn{4}{c}{Non-Linear} \\ \cline{4-7} \cline{9-12}
Pre-int.           & Bootstrap   &  & True ATE  & Bias  & Rel. Bias  & Coverage     &      & True ATE  & Bias  & Rel. Bias  & Coverage \\
4 & block & & 37.12  & 0.33 & 0.009 & 0.93 & & 2.03 & 0.22 & 0.107 & 0.95 \\ 
9 & block & & 316.13 & 1.07 & 0.003 & 0.94 & & 2.07 & 0.15 & 0.074 & 0.94 \\ 
   \hline
\end{tabular}}
\end{table}

\clearpage

\section{Additional application material}
\label{app:application}

\begin{footnotesize}
\begin{longtable}{p{0.2\linewidth}p{0.35\linewidth}p{0.14\linewidth}p{0.2\linewidth}}
\caption{\\ \centering Definition of the variables included in the initial dataset}\\
\label{tabapp:vardef} \\
\hline
\multicolumn{4}{c}{\textbf{Dependent variable}} \\ \hline
\textbf{Variable   name}  & \textbf{Definition}  & \textbf{Time period}  & \textbf{Source}  \\ \hline

Standardized   math test score                                               & Standardized math test score of   fifth-grade students.                                                                                                                                                     & School year 2012/2013-school year   2020/2021 & National Institute for the   Evaluation of Educational Instruction and Training (INVALSI) \\ \hline
\multicolumn{4}{c}{\textbf{Time-invariant   variables}}         \\ \hline
\textbf{Variable   name}  & \textbf{Definition}  & \textbf{Time period}  & \textbf{Source}  \\ \hline
Economic   classification dummies                                            & Without specialization,   non-manufacturing (touristic), non-manufacturing (non-touristic), made in   Italy, other manufacturing                                                                            & 2011                                          & Italian National Institute of   Statistics (Istat)                                        \\
Dummy   district                                                             & LLM with at least one industrial   district                                                                                                                                                                 & 2011                                          & Istat                                                                                     \\
Regional   dummies                                                           & A dummy for each of the 20   Italian regions                                                                                                                                                                &                                               & Istat                                                                                     \\
Population   dummies                                                         & $\leq$ 10,000; (10,000; 50,000{]};   (50,000; 100,000{]}; (100,000; 500,000{]}; \textgreater{}500,000                                                                                                             & 2011                                          & Istat                                                                                     \\
Share of   individuals with a university degree                              & Share of individuals aged 30-34   with a university degree                                                                                                                                                  & 2014                                          & Istat                                                                                     \\
Share of   individuals with a high-school degree                             & Share of individuals aged 30-34   with a high-school degree                                                                                                                                                 & 2014                                          & Istat                                                                                     \\
Share of   urban area                                                        & Urban surface / Total surface                                                                                                                                                                               & 2012                                          & Istat                                                                                     \\
Share of   population in the periphery                                       & Share of population located in   municipalities considered as peripheral or ultra-peripheral according to the   SNAI classification                                                                         & 2011                                          & Istat                                                                                     \\
Number   of public libraries                                                 & Number of public libraries                                                                                                                                                                                  & 2016                                          & Istat                                                                                     \\
Number   of visitors to museums                                              & Number of visitors to museums,   galleries, archaeological sites, and monuments per 100,000 inhabitants                                                                                                     & 2015                                          & Istat                                                                                     \\
Average   maximum temperature                                                & 30-year average (1980-2010) of   maximum air temperature (°C)                                                                                                                                               & 1980-2010                                     & CAMS European Air Quality   Reanalysis Data                                               \\
Average   minimum temperature                                                & 30-year average (1980-2010) of   minimum air temperature (°C)                                                                                                                                               & 1980-2010                                     & CAMS European Air Quality   Reanalysis Data                                               \\
Average   total yearly precipitation                                         & 30-year average (1980-2010) of   total precipitation (mm)                                                                                                                                                   & 1980-2010                                     & CAMS European Air Quality   Reanalysis Data                                               \\
Turnout referendum                                                           & Turnout at the   constitutional referendum that was held in Italy on 4 December 2016                                                                                                                        & 2016                                          & Ministry of the   Interior                                                                \\
Share of   Yes at the referendum                                             & Share of Yes at the   constitutional referendum that was held in Italy on 4 December 2016                                                                                                                   & 2016                                          & Ministry of the Interior                                                                  \\
Percentage   of children aged 0-2 years who have utilized childcare services & Children aged 0-2 years who have   utilized childcare services provided by Municipalities (nurseries,   micro-nurseries, or integrated and innovative services) / Resident children   aged 0-2 years * 100. & 2016                                          & Istat                                                                                     \\ \hline
\multicolumn{4}{c}{\textbf{Time-varying   variables}} \\ \hline
\textbf{Variable   name}  & \textbf{Definition}  & \textbf{Time period}  & \textbf{Source}  \\ \hline
Standardized Italian test score                                              & Standardized   Italian test score of fifth-grade students.                                                                                                                                                  & School year 2012/2013-school year   2018/2019 & INVALSI                                                                                   \\
Percentage   of participants to the math test                                & Percentage of participation in   the math test (pupils who took the test out of expected pupils)                                                                                                            & School year 2012/2013-school year   2018/2019 & INVALSI                                                                                   \\
Percentage   of participants to the Italian test                             & Percentage of participation in   the Italian test (pupils who took the test out of expected pupils)                                                                                                         & School year 2012/2013-school year   2018/2019 & INVALSI                                                                                   \\
Average   percentage score for math                                          & Average percentage score for   math - adjusted for cheating                                                                                                                                                 & School year 2012/2013-school year   2018/2019 & INVALSI                                                                                   \\
Average   percentage score for Italian                                       & Average percentage score for   Italian - adjusted for cheating                                                                                                                                              & School year 2012/2013-school year   2018/2019 & INVALSI                                                                                   \\
Standard   deviation for math                                                & Standard deviation of the   percentage score for math                                                                                                                                                       & School year 2012/2013-school year   2018/2019 & INVALSI                                                                                   \\
Standard   deviation for Italian                                             & Standard deviation of the   percentage score for Italian                                                                                                                                                    & School year 2012/2013-school year   2018/2019 & INVALSI                                                                                   \\
Log of   the Gini index                                                      & Log of the Gini index                                                                                                                                                                                       & 2013-2019                                     & Ministry of Economy and Finance   (MEF)                                                   \\
Share of   foreign population                                                & Foreigners / population                                                                                                                                                                                     & 2013-2019                                     & Istat                                                                                     \\
Unemployment rate                                                            & Resident   population aged 15+ not in employment but currently available for work /   Labor force                                                                                                           & 2013-2019                                     & Istat                                                                                     \\
Share of   graduate mayors                                                   & Share of municipalities with a   mayor with a university degree                                                                                                                                             & 2013-2019                                     & Ministry of the Interior                                                                  \\
Share of   recycled waste                                                    & Share of recycled waste                                                                                                                                                                                     & 2013-2019                                     & Italian National Institute for   Environmental Protection and Research (Ispra)            \\
Share of   workers in manufacturing                                          & Share of workers in   manufacturing                                                                                                                                                                         & 2013-2019                                     & Istat                                                                                     \\
Share of   old population                                                    & Share of population aged \textgreater{}=65                                                                                                                                                                  & 2013-2019                                     & Istat                                                                                     \\
Declared   income per capita                                                 & Declared income per capita                                                                                                                                                                                  & 2013-2019                                     & MEF                                                                                       \\
Log of   declared income (total)                                             & Log of declared income (total)                                                                                                                                                                              & 2013-2019                                     & MEF                                                                                       \\
Share of   income for pensions                                               & Share of overall declared income   for pensions                                                                                                                                                             & 2013-2019                                     & MEF                                                                                       \\
Share of   income for employment                                             & Share of overall declared income   for employment                                                                                                                                                           & 2013-2019                                     & MEF                                                                                       \\
Population                                                                   & Resident population                                                                                                                                                                                         & 2013-2019                                     & Istat                                                                                     \\
Average   price per square meter                                             & Average price per square meter -   house                                                                                                                                                                    & 2013-2019                                     & Osservatorio del Mercato   Immobiliare – Agenzia delle Entrate   \\ \hline
\multicolumn{4}{p{1\textwidth}}{\footnotesize \textit{Notes.} In our application, time-invariant covariates are either features of the LLM not subject to time changes (e.g. area of the LLM) or features taken at a time before 2017. Data from the 2020/2021 school year represent the first available post-pandemic data point for the standardized math test score, as there was no test for the school year 2019/2020 due to the national lockdown. Following \citet{Carlana:Ferrara:Lopez:2023}, math test scores have been standardized for the entire dataset using the 2013-2019 values as a benchmark, with a mean of 0 and a standard deviation of 1.}
\end{longtable}

\end{footnotesize}

\clearpage
\begin{table}[H]
\centering
\caption{Performances with different numbers of variables
}
\label{tabapp:performance}
\begin{tabular}{lccc} \hline
\textbf{Method} & \multicolumn{3}{c}{\textbf{MSE}} \\ \hline
& \textbf{10 variables} & \textbf{20 variables} & \textbf{30 variables} \\ \cline{2-4}
LASSO                            & 0.4316                & 0.4169                & 0.6559                \\
Partial Least Squares            & 0.4172                & 0.4044                & 0.4156                \\
Boosting                         & 0.3995                & 0.3909                & 0.4037                \\
Random Forest                    & 0.4011                & 0.3902                & 0.3935  \\ \hline             
\end{tabular}
\end{table}

\clearpage

\begin{table}[H]
\caption{Definition of the 20 variables included in the estimation process (random forest)}
\label{tabapp:rf_included_vars}
\begin{tabular}{ll}
\hline
\textbf{Predictors}                                  & \textbf{Lag}            \\ \hline
Standardized   math test score                       & First, second and third \\
Standardized   Italian test score                    & First, second and third \\
Average   percentage score for math                  & First, second and third \\
Change   in the average percentage score for Italian & First                   \\
Average   percentage score for Italian               & Second                  \\
Change   in the average percentage score for Italian & First                   \\
Change   in the standard deviation for math          & First and second        \\
Change   in the unemployment rate                    & Third                   \\
Change   in the log of declared income (total)       & Third                   \\
Log of   declared income (total)                     & Third                   \\
Share of   individuals with a high-school degree     & Fixed (2014)            \\
Dummy   Calabria                                     & Fixed                   \\
Turnout   referendum                                 & Fixed (2016)           \\
\hline
\multicolumn{2}{p{0.9\textwidth}}{\footnotesize \textit{Note}: These are the 20 variables selected via the panel CV procedure by the preliminary random forest.}
\end{tabular}
\end{table}

\clearpage

\begin{footnotesize}
\begin{longtable}{p{0.2\linewidth}p{0.35\linewidth}p{0.15\linewidth}p{0.2\linewidth}} 
\caption{\\ \centering Definition of the variables included in the data-driven CATEs analysis}\\
\label{tabapp:cate_vars} \\
\hline
\textbf{Variable   name}                                                   & \textbf{Definition}                                                                                                                                                              & \textbf{Time period}              & \textbf{Source}                                                \\ \hline
COVID-19   impact on the standardized math test score (dependent variable) & Treatment effect of the COVID-19   crisis on the standardized math test score (SD)                                                                                               & School year 2020/2021             & Estimated via the MLCM with random   forest                    \\
Share of   income accruing to low-income earners                           & Share of income accruing to   individuals with an overall income $\leq$ €26,000                                                                                                        & 2019                              & MEF                                                            \\
Average   download speed of the internet            & Average download speed of the   internet                                                                                                                  & 2018                              & AGCOM                                                          \\
Share   of households without internet access                              & Share of households without   internet access                                                                                                                                    & 2018                              & AGCOM                                                          \\
Tourism   LLM                                                              & Dummy variable equal to 1 for   LLMs specialized in tourism                                                                                                                      & 2011                              & Istat                                                          \\
Share of   individuals with a university degree                            & Share of individuals aged 30-34   with a university degree                                                                                                                       & 2015                              & Istat                                                          \\
Population                                                                 & Resident population                                                                                                                                                              & 2019                              & Istat                                                          \\
Per   capita expenditure on gambling                                       & Per capita expenditure on   gambling                                                                                                                                             & 2019                              &                                                                \\
Unemployment   rate                                                        & Resident population aged 15+ not   in employment but currently available for work                                                                                                & 2019                              & Istat                                                          \\
Excess   mortality estimates                                               & Municipality-level excess   mortality estimated by applying ML techniques to all-cause deaths data,   aggregated at the LLM level                                                & From Feb 21, 2020 to Sep 30, 2020 & \citet{Cerqua:DiStefano:Letta:Miccoli:2021}                                           \\
Share of   temporary contracts                                             & Number of employees with   temporary contracts in October divided by the number of employees in October                                                                          & 2015                              & Istat                                                          \\
Share of   jobs in suspended economic activities                           & Share of jobs in activities   suspended in March 2020 by the Italian Government due to the spread of the   pandemic                                                              & 2017                              & Istat                                                          \\
Per   capita income                                                        & The amount of money earned per   person                                                                                                                                          & 2019                              & MEF                                                            \\
Share   of firms having employees in short-time work subsidies             & The number of firms with   employees receiving short-time work subsidies divided by the universe of   firms registered in the Business Register                                  & Average (2015-2018)               & Ministry of Labor and Social   Policies                        \\
Share of   population living in peripheral areas                           & Share of population living in   areas defined by Istat as peripheral or ultra-peripheral                                                                                         & Jan 1, 2020                       & Istat                                                          \\
Dependency ratio                                                           & The ratio of   those typically not in the labor force (the dependent part, ages 0 to 14 and   65+) and those typically in the labor force (the productive part, ages 15 to   64) & Jan 1, 2020                       & Istat                                                          \\
Gini   index                                                               & Gini index                                                                                                                                                                       & 2019                              & MEF                                                            \\
Average   price per square meter                                           & Average price per square meter -   house                                                                                                                                         & 2019                              & Osservatorio del Mercato   Immobiliare – Agenzia delle Entrate \\ \hline
\end{longtable}
\end{footnotesize}

\newpage
\begin{figure}[H]
    \centering
    \includegraphics[scale = 0.6]{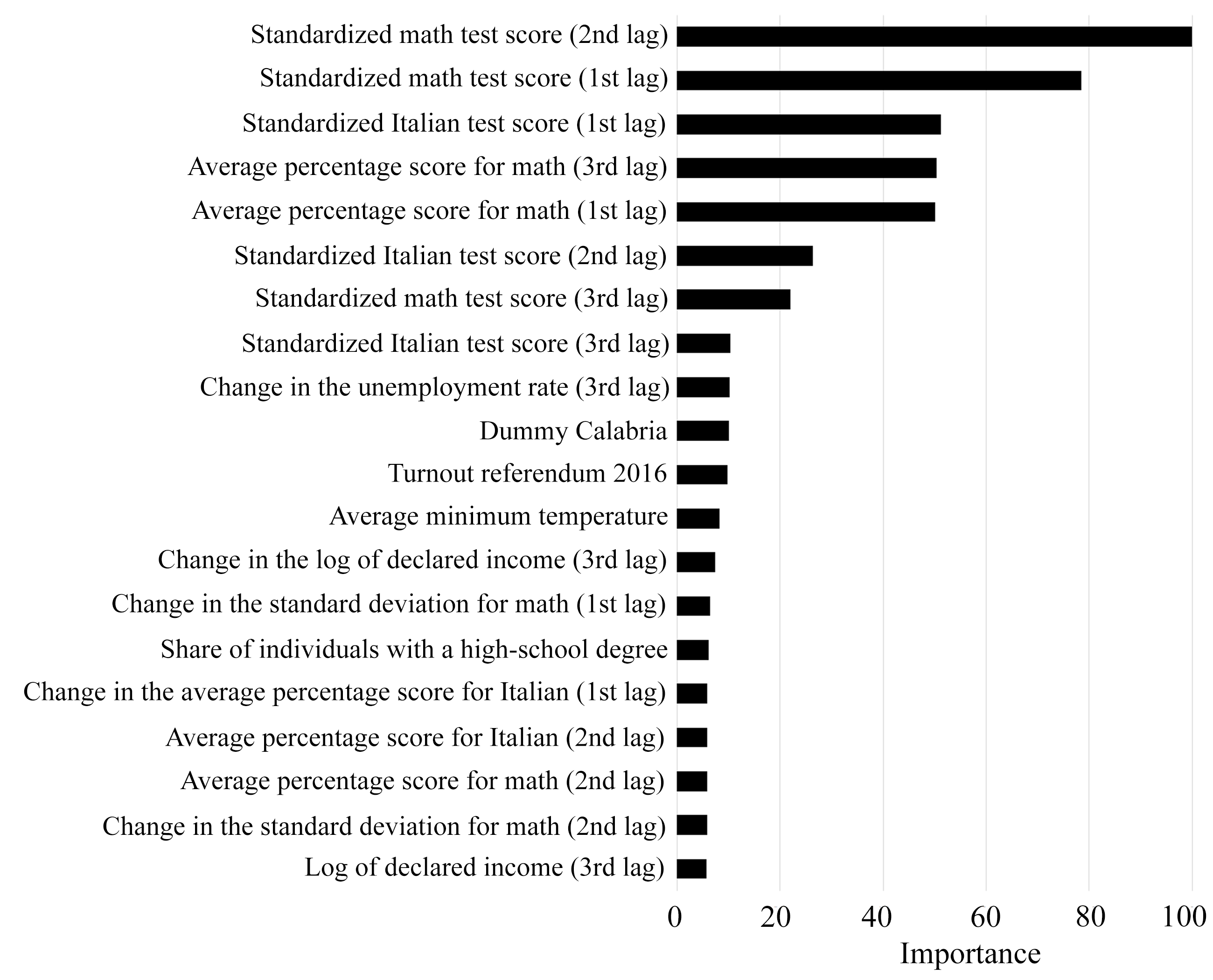}
    \caption{Variable importance ranking – Preliminary random forest.}
    \label{appfig:rf_varimp}
\end{figure}

\begin{figure}[H]
    \centering
    \includegraphics[scale = 0.3]{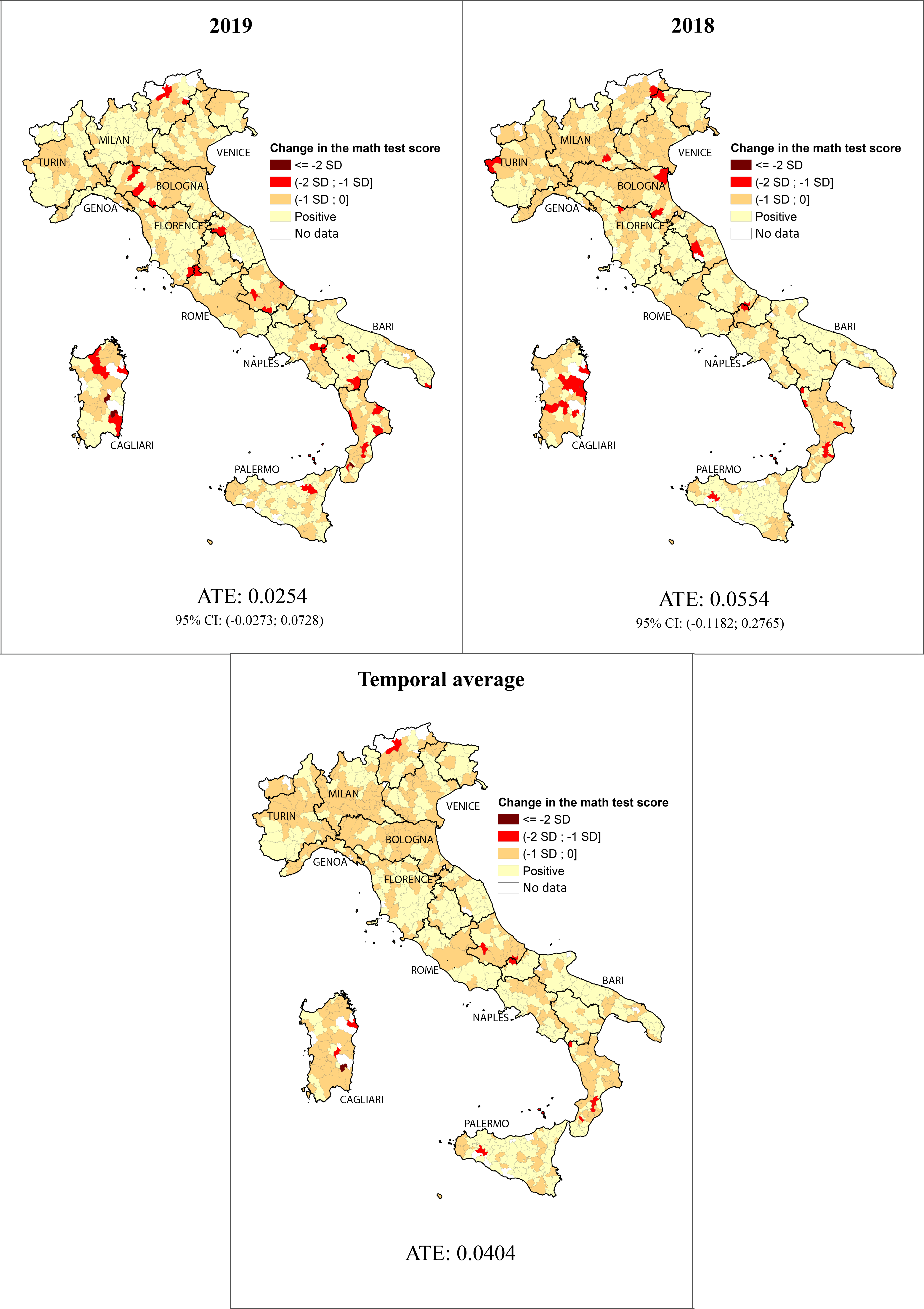}
    \caption{Unit-level panel placebo test by year.}
    \label{appfig:yearly_placebos}
\end{figure}

\newpage
\begin{figure}[H]
    \centering
    \includegraphics[scale = 0.95]{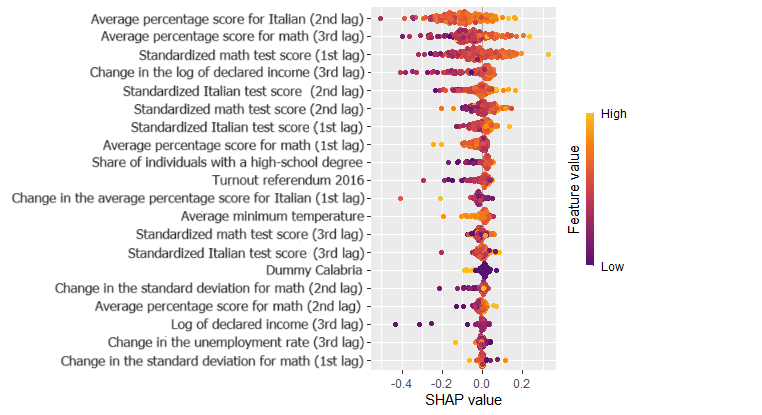}
    \caption{SHAP values of the counterfactual 2020 forecasts produced by the final selected model (random forest) for a random 20\% subsample of LLMs.}
    \label{appfig:shapley}
\end{figure}

\clearpage

\section{Replication study}
\label{app:replication}
In this section, we revisit the empirical application of \citet{callaway2021difference} on the impact of minimum wage policy on teen employment in the United States (US). The main goal of this exercise is to assess whether the MLCM can replicate findings from a study that used a popular method based on the canonical use of control units. Additionally, we demonstrate how the MLCM can be easily adapted to a setting with staggered adoption of treatment. For the purpose of this replication, we simulate an environment in which the researcher has access to but cannot employ untreated units due to violations of the no-interference assumption. We check the reliability of these estimates by using the results of \citet{callaway2021difference}, obtained via their difference-in-differences estimator for staggered adoption contexts with multiple time periods, as the ground truth.

We use county-level data for the period 2001-2007 from \citet{callaway2021difference} and apply only minimal pre-processing before proceeding with estimation. First, we discard never-treated units. Second, we employ the same set of variables included by \citet{callaway2021difference} in their specification with covariates: region dummies, county population, county median income, the fraction of white population that is white, the fraction of the population with a high school education, the county’s poverty rate. Third, we also include, as additional predictors, a lag of the outcome variable (the log of teen employment), county fixed effects---using the approach recommended by \citet{Johannemann:Hadad:Athey:Wager:2019}---and region-year dummies.

We then proceed to estimate the effects of staggered implementation of minimum wage policy on teen employment by forecasting counterfactuals with the MLCM. The key point to keep in mind is that not only we discard never-treated units, but we also do not rely on the available not-yet-treated units to build counterfactuals. Regarding the identification assumptions, we assume additivity and no anticipation, as in \citet{callaway2021difference}, but we replace their parallel trends assumption with our Assumption \ref{assumpt:model} that implicitly rules out concomitant shocks and policies. This assumption appears reasonable given the context: the years under analysis are those just before the Great Recession, during which no other major employment shock occurred.\footnote{Youth unemployment trends remained relatively stable in the US during these years, with a consistent value of 10.5\% in both 2001 and 2007 (the initial and final years of the dataset); see data from the World Bank \href{https://data.worldbank.org/indicator/SL.UEM.1524.ZS?locations=US}{here}.} We also replace their standard SUTVA assumption, which postulates no interference among units, with our weaker version of SUTVA (Assumption \ref{assumpt:sutva}), which only requires no hidden versions of the treatment. 

We employ two versions of the MLCM: one with a fully non-linear model, random forest, and the other with a linear model, LASSO. LASSO outperformed random forest on this specific dataset across all treatment cohorts, making it the chosen method for forecasting counterfactual teen employment levels in the post-treatment periods.\footnote{For this reason, we do not need to rely on Assumption \ref{assumpt:additional_models}, which is only required for identification in \textit{non-linear} multi-step ahead forecasting.}

The results of this no-control replication exercise are reported in the event-study graph of Figure \ref{appfig:cs} below. This graph illustrates group-time average treatment effects (i.e., the estimand proposed by \citet{callaway2021difference}) and should be compared with the two sets of estimates reported in Figure 1 of the original study. 

\begin{figure}[H]
    \centering
    \includegraphics[scale = 0.78]{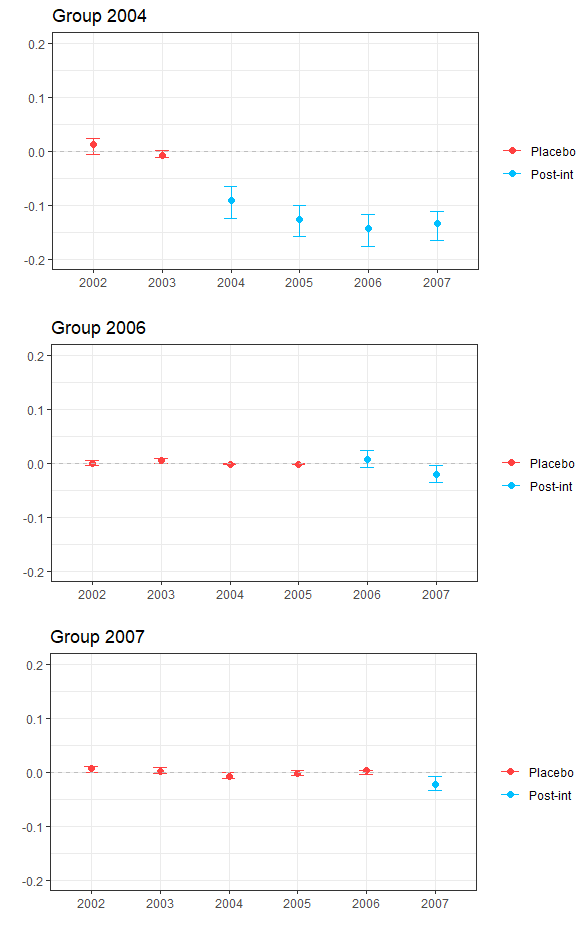}
    \caption{Group-time average treatment effects -- Replication without controls of \citet{callaway2021difference}. \textit{Note}: Confidence intervals estimated using 1000 block-bootstrap replications. To ease comparability, the range of the Y axis is the same used by \citet{callaway2021difference}.}
    \label{appfig:cs}
\end{figure}
\vspace{-0.5cm}
Despite completely discarding information from untreated units, we closely replicate their findings: the sign of the post-treatment estimates is always consistent (with the sole exception of the 2006 effect for the cohort first treated in 2006), while the magnitude differs to some extent but consistently falls within the range of confidence intervals from the two specifications of \citet{callaway2021difference} with conditional and unconditional parallel trends. The qualitative story remains unchanged, pointing to negative impacts of the minimum wage increase on county-level teen employment: our estimate of the `global' average treatment effect is –3\% (95\% confidence intervals: –3.8;  –2), which aligns with the estimate reported by \citet{callaway2021difference} of –3.2\% for their specification with conditional parallel trends.\footnote{\citet{callaway2021difference} refer to this estimate as the `simple weighted average'. It is the weighted average (by group size) of all available group-time average treatment effects.} Lastly, pre-treatment estimates are all indistinguishable from zero. This is an improvement over the original study, where some of the placebo estimates were relatively large and statistically significant, casting doubts on the validity of the parallel trends assumption \citep{callaway2021difference}.

\begin{figure}[H]
    \centering
    \includegraphics[scale = 0.52]{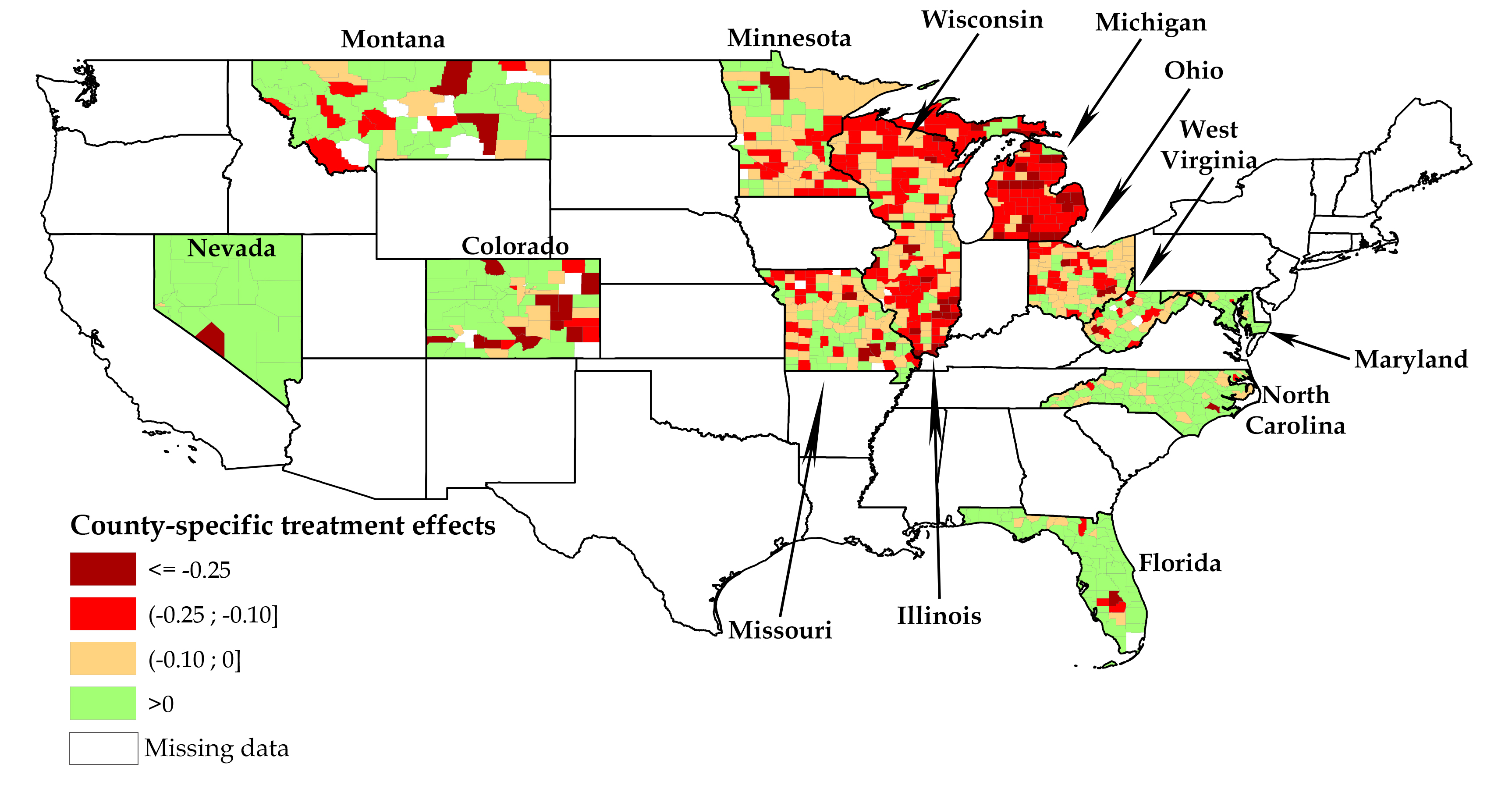}
    \centering{\caption{County-specific treatment effects of the minimum wage on teen employment. \textit{Notes}: These estimates are individual treatment effects averaged over the post-treatment period. The final sample used by \citet{callaway2021difference} includes county-level teen employment data for 29 states, matched with county characteristics. Of the 29 states considered in the \citet{callaway2021difference} analysis, 13 are treated. These 13 states comprise 930 counties. However, due to missing data in the original dataset, the number of treated counties is 907.}
    \label{appfig:map_cs}}
\end{figure}

Moreover, our county-specific treatment effects can be exploited to investigate patterns of treatment effect heterogeneity. One approach is to leverage these unit-level estimates by mapping them to explore potential spatial patterns. Figure \ref{appfig:map_cs} shows county-specific effects, revealing highly heterogeneous estimates across states. Notably, the minimum wage had a particularly strong negative impact in Michigan, which is also the state that increased the minimum wage the most during the period under analysis \citep{callaway2021difference}.

This replication study demonstrates the flexibility and applicability of the MLCM in staggered adoption settings or in cases with potential spillovers and general equilibrium effects contaminating the outcomes of untreated units. It also highlights that the MLCM can be used as a robustness check for traditional identification strategies, offering an alternative estimation method to assess the plausibility of the key identification assumptions, validate the main estimates, and rule out spillover effects. Finally, the MLCM generates individual treatment effects, providing a more granular building block compared to the group-time average effects of \citet{callaway2021difference} and most other traditional estimators leveraging control units, thus allowing for a more comprehensive investigation into the heterogeneity of impacts.
\end{appendices}

\end{document}